%% file: main.tex
\documentclass[a4paper,11pt]{article}
\pdfoutput=1 

\usepackage{jcappub} 
\usepackage{bm}
\usepackage[table,xcdraw]{xcolor}
\usepackage{hyperref}
\usepackage{graphicx}
\usepackage[noabbrev]{cleveref}
\usepackage[T1]{fontenc} 
\usepackage{longtable}
\usepackage[tableposition=t]{caption}
\usepackage{booktabs}
\usepackage{rotating}
\usepackage{enumitem}
\usepackage{float}
\usepackage{orcidlink}
\usepackage{nicefrac}
\usepackage[normalem]{ulem}
\input{mydefinitions}

\usepackage{amsmath}
\let\oldAA\AA
\renewcommand{\AA}{\text{\normalfont\oldAA}}

\title{Synthetic spectra for Lyman-$\alpha$ forest analysis in the Dark Energy Spectroscopic Instrument.}

\author[1]{{Hiram~K.~Herrera-Alcantar}\orcidlink{0000-0002-9136-9609},}
\author[2,a]{{Andrea~Muñoz-Gutiérrez}\orcidlink{0000-0002-3565-0148},\note[a]{Corresponding author.}}
\author[3,4]{{Ting~Tan}\orcidlink{0000-0001-8289-1481},}
\author[1,5]{{Alma~X.~González-Morales}\orcidlink{0000-0003-4089-6924},}
\author[6]{{Andreu~Font-Ribera}\orcidlink{0000-0002-3033-7312},}
\author[7]{{Julien~Guy}\orcidlink{0000-0001-9822-6793},}
\author[8]{{John~Moustakas}\orcidlink{0000-0002-2733-4559},}
\author[9]{{David~Kirkby}\orcidlink{0000-0002-8828-5463},}

\author[10]{{E.~Armengaud}\orcidlink{0000-0001-7600-5148},}
\author[11]{{A.~Bault}\orcidlink{0000-0002-9964-1005},}
\author[6]{{L.~Cabayol-Garcia},}
\author[6]{{J.~Chaves-Montero}\orcidlink{0000-0002-9553-4261},}
\author[12,b]{{A.~Cuceu}\orcidlink{0000-0002-2169-0595},\note[b]{Nasa Einstein Fellow.}}
\author[1]{{R.~de la Cruz}\orcidlink{0000-0001-9908-9129},}
\author[13]{{L.~\'A.~Garc\'ia}\orcidlink{0000-0003-1235-794X},}
\author[6]{{C.~Gordon},}
\author[14]{{V.~Ir\v{s}i\v{c}}\orcidlink{0000-0002-5445-461X},}
\author[12,15,16,17]{{N.~G.~Kara{\c c}ayl{\i}}\orcidlink{0000-0001-7336-8912},}
\author[10]{{J.M.~Le~Goff},}
\author[18]{{P.~Montero-Camacho}\orcidlink{0000-0002-6998-6678},}
\author[1,19]{{G.~Niz}\orcidlink{0000-0002-1544-8946},}
\author[20]{{I.~P\'erez-R\`afols}\orcidlink{0000-0001-6979-0125},}
\author[6]{{C.~Ram\'irez-P\'erez},}
\author[10,21]{{C.~Ravoux}\orcidlink{0000-0002-3500-6635},}
\author[22,23]{{M.~Walther}\orcidlink{0000-0002-1748-3745},}
\author[24]{{J.~Aguilar},}
\author[25]{{S.~Ahlen}\orcidlink{0000-0001-6098-7247},}
\author[26]{{D.~Brooks},}
\author[24]{{T.~Claybaugh},}
\author[27]{{K.~Dawson},}
\author[2]{{A.~de la Macorra}\orcidlink{0000-0002-1769-1640},}
\author[26]{{P.~Doel},}
\author[28,29]{{J.~E.~Forero-Romero}\orcidlink{0000-0002-2890-3725},}
\author[30,31,32]{{E.~Gaztañaga},}
\author[24]{{S.~Gontcho A Gontcho}\orcidlink{0000-0003-3142-233X},}
\author[12,16,17]{{K.~Honscheid},}
\author[33]{{R.~Kehoe},}
\author[24]{{T.~Kisner}\orcidlink{0000-0003-3510-7134},}
\author[24]{{M.~Landriau}\orcidlink{0000-0003-1838-8528},}
\author[24]{{Michael~E.~Levi}\orcidlink{0000-0003-1887-1018},}
\author[6,34]{{M.~Manera}\orcidlink{0000-0003-4962-8934},}
\author[12,15,17]{{P.~Martini}\orcidlink{0000-0002-4279-4182},}
\author[35]{{A.~Meisner}\orcidlink{0000-0002-1125-7384},}
\author[6,36]{{R.~Miquel},}
\author[37]{{J.~Nie}\orcidlink{0000-0001-6590-8122},}
\author[10,24]{{N.~Palanque-Delabrouille}\orcidlink{0000-0003-3188-784X},}
\author[24,38,39]{{C.~Poppett},}
\author[40]{{M.~Rezaie}\orcidlink{0000-0001-5589-7116},}
\author[41]{{G.~Rossi},}
\author[42]{{E.~Sanchez}\orcidlink{0000-0002-9646-8198},}
\author[43]{{H.~Seo}\orcidlink{0000-0002-6588-3508},}
\author[44]{{G.~Tarl\'{e}}\orcidlink{0000-0003-1704-0781},}
\author[35]{{B.~A.~Weaver},}
\author[37]{{Z.~Zhou}\orcidlink{0000-0002-4135-0977},}

\affiliation[1]{Departamento de Física, División de Ciencias e Ingenierías, Campus León, Universidad de Guanajuato, 37150, León, Guanajuato, México.}
\affiliation[2]{Instituto de Física, Universidad Nacional Autónoma de México, Apdo. Postal 20-364, México}
\affiliation[3]{Sorbonne Universit\'e, CNRS/IN2P3, Laboratoire de Physique Nucl\'eaire et de Hautes Energies, LPNHE, 4 Place Jussieu, F-75252 Paris, France}
\affiliation[4]{CNRS-UCB International Research Laboratory, Centre Pierre Binétruy, IRL2007, CPB-IN2P3, Berkeley, US}
\affiliation[5]{Consejo Nacional de Ciencia y Tecnología, Av. Insurgentes Sur 1582. Colonia Credito Constructor, Del. Benito Juarez C.P. 03940, México D.F. México}
\affiliation[6]{Institut de Física d’Altes Energies (IFAE), The Barcelona Institute of Science and Technology, Campus UAB, 08193 Bellaterra Barcelona, Spain}
\affiliation[7]{Lawrence Berkeley National Laboratory, 1 Cyclotron Road, Berkeley, CA 94720, USA}
\affiliation[8]{Department of Physics and Astronomy, Siena College, 515 Loudon Road, Loudonville, NY 12211, USA}
\affiliation[9]{Department of Physics and Astronomy, University of California, Irvine, 92697, USA}

\affiliation[10]{IRFU, CEA, Universit\'{e} Paris-Saclay, F-91191 Gif-sur-Yvette, France}
\affiliation[11]{Department of Physics and Astronomy, University of California, Irvine, 92697, USA}
\affiliation[12]{Center for Cosmology and AstroParticle Physics, The Ohio State University, 191 West Woodruff Avenue, Columbus, OH 43210, USA}
\affiliation[13]{Universidad ECCI, Cra. 19 No. 49-20, Bogot\'a, Colombia, C\'odigo Postal 111311}
\affiliation[14]{Kavli Institute for Cosmology, University of Cambridge, Madingley Road, Cambridge CB3 0HA, UK}
\affiliation[15]{Department of Astronomy, The Ohio State University, 4055 McPherson Laboratory, 140 W 18th Avenue, Columbus, OH 43210, USA}
\affiliation[16]{Department of Physics, The Ohio State University, 191 West Woodruff Avenue, Columbus, OH 43210, USA}
\affiliation[17]{The Ohio State University, Columbus, 43210 OH, USA}
\affiliation[18]{Department of Astronomy, Tsinghua University, 30 Shuangqing Road, Haidian District, Beijing, China, 100190}
\affiliation[19]{Instituto Avanzado de Cosmolog\'{\i}a A.~C., San Marcos 11 - Atenas 202. Magdalena Contreras, 10720. Ciudad de M\'{e}xico, M\'{e}xico}
\affiliation[20]{Departament de F\'isica, EEBE, Universitat Polit\`ecnica de Catalunya, c/Eduard Maristany 10, 08930 Barcelona, Spain}
\affiliation[21]{Aix Marseille Univ, CNRS/IN2P3, CPPM, Marseille, France}
\affiliation[22]{Excellence Cluster ORIGINS, Boltzmannstrasse 2, D-85748 Garching, Germany}
\affiliation[23]{University Observatory, Faculty of Physics, Ludwig-Maximilians-Universit\"{a}t, Scheinerstr. 1, 81677 M\"{u}nchen, Germany}
\affiliation[24]{Lawrence Berkeley National Laboratory, 1 Cyclotron Road, Berkeley, CA 94720, USA}
\affiliation[25]{Physics Dept., Boston University, 590 Commonwealth Avenue, Boston, MA 02215, USA}
\affiliation[26]{Department of Physics \& Astronomy, University College London, Gower Street, London, WC1E 6BT, UK}
\affiliation[27]{Department of Physics and Astronomy, The University of Utah, 115 South 1400 East, Salt Lake City, UT 84112, USA}
\affiliation[28]{Departamento de F\'isica, Universidad de los Andes, Cra. 1 No. 18A-10, Edificio Ip, CP 111711, Bogot\'a, Colombia}
\affiliation[29]{Observatorio Astron\'omico, Universidad de los Andes, Cra. 1 No. 18A-10, Edificio H, CP 111711 Bogot\'a, Colombia}
\affiliation[30]{Institut d'Estudis Espacials de Catalunya (IEEC), 08034 Barcelona, Spain}
\affiliation[31]{Institute of Cosmology and Gravitation, University of Portsmouth, Dennis Sciama Building, Portsmouth, PO1 3FX, UK}
\affiliation[32]{Institute of Space Sciences, ICE-CSIC, Campus UAB, Carrer de Can Magrans s/n, 08913 Bellaterra, Barcelona, Spain}
\affiliation[33]{Department of Physics, Southern Methodist University, 3215 Daniel Avenue, Dallas, TX 75275, USA}
\affiliation[34]{Departament de F\'{i}sica, Serra H\'{u}nter, Universitat Aut\`{o}noma de Barcelona, 08193 Bellaterra (Barcelona), Spain}
\affiliation[35]{NSF's NOIRLab, 950 N. Cherry Ave., Tucson, AZ 85719, USA}
\affiliation[36]{Instituci\'{o} Catalana de Recerca i Estudis Avan\c{c}ats, Passeig de Llu\'{\i}s Companys, 23, 08010 Barcelona, Spain}
\affiliation[37]{National Astronomical Observatories, Chinese Academy of Sciences, A20 Datun Rd., Chaoyang District, Beijing, 100012, P.R. China}
\affiliation[38]{Space Sciences Laboratory, University of California, Berkeley, 7 Gauss Way, Berkeley, CA  94720, USA}
\affiliation[39]{University of California, Berkeley, 110 Sproul Hall \#5800 Berkeley, CA 94720, USA}
\affiliation[40]{Department of Physics, Kansas State University, 116 Cardwell Hall, Manhattan, KS 66506, USA}
\affiliation[41]{Department of Physics and Astronomy, Sejong University, Seoul, 143-747, Korea}
\affiliation[42]{CIEMAT, Avenida Complutense 40, E-28040 Madrid, Spain}
\affiliation[43]{Department of Physics \& Astronomy, Ohio University, Athens, OH 45701, USA}
\affiliation[44]{University of Michigan, Ann Arbor, MI 48109, USA}

\emailAdd{andreamgtz@ciencias.unam.mx}

\abstract{
 Synthetic data sets are used in cosmology to test analysis procedures, to verify that systematic errors are well understood and to demonstrate that measurements are unbiased. In this work we describe the methods used to generate synthetic datasets of Lyman-$\alpha$ quasar spectra aimed for studies with the Dark Energy Spectroscopic Instrument (DESI). In particular, we focus on demonstrating that our simulations reproduces important features of real samples, making them suitable to test the analysis methods to be used in DESI and to place limits on systematic effects on measurements of Baryon Acoustic Oscillations (BAO). 
 We present a set of mocks that reproduce the statistical properties of the DESI early data set with good agreement. Additionally, we use a synthetic dataset to forecast the BAO scale constraining power of the completed DESI survey through the Lyman-$\alpha$ forest.}

\begin{document}
\maketitle
\flushbottom

\input{introduction}
\input{mockspectra}

\input{earlymocks}

\input{Y5mocks}

\section{Summary and Conclusions}
We have presented the methodology to produce synthetic DESI Lyman-$\alpha$ datasets. This methodology implements the use of the script \texttt{quickquasars}, which compiles multiple modules from the \texttt{desisim} and \texttt{specsim} repositories of the DESI main code dedicated to produce synthetic Lyman-$\alpha$ spectra, and has the following characteristics:
\begin{itemize}[noitemsep]
    \item It requires as input the raw transmitted flux in order to produce noiseless Lyman-$\alpha$ spectra by combining the raw flux with a quasar-continuum template.
    \item It can reproduce a survey given its footprint, object number density distribution and redshift-magnitude distributions. This feature supports not only full DESI-Y5 mocks, but also early stages of the survey, and even other surveys with known characteristics, including future projects such as DESI-II.
    \item Instrumental noise is added to spectra given an instrument response model and observing conditions. Models for instruments other than DESI can be used if available. 
    \item  Astrophysical contaminants can be added to spectra. The types of contaminants available are DLAs, BALs and metals. This feature is important to study the effect of each type of contaminant in the measurement of the Lyman-$\alpha$ forest correlation functions.
\end{itemize}
With this methodology we were able to produce 40 mock datasets that qualitatively reproduce the observed DESI EDR+M2 dataset, 20 for \texttt{LyaCoLoRe} and 20 for the \texttt{Saclay} raw transmissions files used as input including: 10 mocks containing only Lyman-$\alpha$ absorption only and 10 mocks additionally including DLAs, BALs and metals. As discussed in \cref{sec:mockvsdata}, these mocks qualitatively reproduce the statistical properties of observed data. Regarding the correlation functions, the dispersion on the measurements of our mock realizations is enough to be overall consistent with the measured correlation functions on observed data with differences that where discussed in \cref{subsec:corrfuncs_EDR}.

The favorable results obtained do not suggest a need for major changes in our mock generation methodology, except for the sampling method used to mimic the EDR+M2 footprint and quasar number density discussed during \cref{subsec:survey_attribs}. Nevertheless, updates may be considered for future DESI releases if required. For example: using the observed DESI data to improve the instrumental noise model to better reproduce the observations for bright quasars, produce and include new sets of BAL templates, updating the DLA $f(N_{\rm{HI}})$ distribution, re-tuning the coefficients used to add metals to the spectra. The last of these point might also require exploring a new method following a non-linear relationship between metals and Lyman-$\alpha$.

As the qualitative comparison of our EDR+M2 mocks results with those of observed data was favorable, we made DESI-Y5 mocks including only Lyman-$\alpha$ absorption and performed a forecast in the constraining power of the BAO scale parameters $\alpha_\parallel$ and $\alpha_\perp$. Our results are consistent with the Fisher forecast performed in \citep{DESI:2016fyo}. Forecasts using a more realistic dataset including contaminants could be done to test the effect of the different parameter models in the constraining power of DESI, however doing so requires further study.

It is important to mention that mocks produced with \texttt{quickquasars} have been used to test systematics in the analysis pipelines and as validation steps in other works before. The eBOSS DR16~\citep{2020duMasDR16}  BAO study with Lyman-$\alpha$ forest included the use of eBOSS mocks as a validation step, these same mocks were used to test the efficiency and purity of the eBOSS DR16 DLA catalog~\citep{chabanier2022completed}. DESI mocks have been used to study the effect of quasar redshift errors on Lyman-$\alpha$ forest correlation functions~\citep{Youles:2022pgf}, test the efficiency and purity of the BAL~\citep{Filbert:2023inprep} and DLA~\citep{wang2022deep} finder algorithms, study the impact of BALs on quasar redshift measurements~\citep{Garcia:2023ddw}, and for validation of P1D studies~\citep{Ravoux:2023pgi,Karacayli:2023afs}. 

The EDR+M2 mocks produced in this work are being used to validate the analysis on the observed EDR+M2 dataset, for example studying the impact of QSO redshift errors in the Lyman-$\alpha$ -- QSO cross-correlation function~\citep{Bault:2023inprep} and to characterize instrumental effects on the measurement of the correlation functions~\citep{Guy:2023inprep}. Future mocks including more realistic features may be produced for other studies during the DESI-Y1 phase. 

\section*{Data Points}
The data points corresponding to each figure in this paper can be accessed in the Zenodo repository at \url{https://doi.org/10.5281/zenodo.10433340}.
\appendix

\acknowledgments
The authors thank Ian McGreer for his contributions and efforts to integrate \texttt{simqso} into the \texttt{desisim} package and consequently into \texttt{quickquasars}. Special thanks to James Rich and Julian Bautista for their invaluable comments and discussions on this manuscript. The authors would also like to thank James Farr, Hélion du Mas des Bourboux, Thomas Etourneau, and all the persons involved on the production of the \texttt{LyaCoLoRe} and \texttt{Saclay} mocks.

HKHA and AXGM acknowledge support from Dirección de Apoyo a la Investigación y al Posgrado, Universidad de Guanajuato, research Grant No. 179/2023 and CONACyT México under Grants No. 286897 and A1-S-17899. 

AFR acknowledges financial support from the Spanish Ministry of Science and Innovation under the Ramon y Cajal program (RYC-2018-025210) and the PGC2021-123012NB-C41 project, and from the European Union's Horizon Europe research and innovation programme (COSMO-LYA, grant agreement 101044612). IFAE is partially funded by the CERCA program of the Generalitat de Catalunya

This material is based upon work supported by the U.S. Department of Energy (DOE), Office of Science, Office of High-Energy Physics, under Contract No. DE–AC02–05CH11231, and by the National Energy Research Scientific Computing Center, a DOE Office of Science User Facility under the same contract. Additional support for DESI was provided by the U.S. National Science Foundation (NSF), Division of Astronomical Sciences under Contract No. AST-0950945 to the NSF’s National Optical-Infrared Astronomy Research Laboratory; the Science and Technology Facilities Council of the United Kingdom; the Gordon and Betty Moore Foundation; the Heising-Simons Foundation; the French Alternative Energies and Atomic Energy Commission (CEA); the National Council of Humanities, Science and Technology of Mexico (CONAHCYT); the Ministry of Science and Innovation of Spain (MICINN), and by the DESI Member Institutions: \url{https://www.desi.lbl.gov/collaborating-institutions}. Any opinions, findings, and conclusions or recommendations expressed in this material are those of the author(s) and do not necessarily reflect the views of the U. S. National Science Foundation, the U. S. Department of Energy, or any of the listed funding agencies.

The authors are honored to be permitted to conduct scientific research on Iolkam Du’ag (Kitt Peak), a mountain with particular significance to the Tohono O’odham Nation.

\numberwithin{table}{section}
\numberwithin{figure}{section}
\input{appendix_continuum}
\input{appendix_rawmockformat}

\input{appendix_output}

\bibliographystyle{JHEP.bst}
\bibliography{references}

\end{document}

%% file: mydefinitions.tex
\newcommand{\lya}{Lyman-$\alpha$ }


%% file: introduction.tex
\section{Introduction}\label{sec:introduction}
The Lyman-$\alpha$ forest is a series of absorption features present in quasar\footnote{Also referred indistinctly as quasi-stellar object or QSO.} spectra caused by intervening neutral hydrogen (HI) clouds along its line of sight and has been proven to be a powerful cosmological tool due to the tight relation between the Lyman-$\alpha$ optical depth and   the densities of gas and dark matter. 
Lyman-$\alpha$ absorption observed in spectra of distant quasars is now widely used to: (i) calculate 3D correlation functions to study Baryon Acoustic Oscillations (BAO)~\citep[e.g.][]{2011SlosarBOSS,2013BuscaBOSS,2013SlosarDR9,Font2014QsoxLYA,2015DelubacDR11,2017BautistaDR12,2017duMasDR12,2019deSainteAgatheDR14,2019BlomqvistDR14,2020duMasDR16}; (ii) measure the line-of-sight one-dimensional flux power spectrum ($P_{\rm{1D}}$) to constrain the amplitude and shape of the matter power spectrum at high redshifts~\citep[e.g.][]{Croft:1997jf,2002Croft:P1D, McDonaldSDSS:2004kjl, 2004KimP1D,2017IrsicP1D, Walther:2017cir,Chabanier:2018rga, 2022KaracayliP1D, Karacayli:2023afs, Ravoux:2023pgi}, and thus to yield tight constraints on neutrino 
masses~\citep[e.g.][]{Palanque-Delabrouille:2015pga,Yeche:2017upn,Palanque-Delabrouille:2019iyz} and dark matter models~\citep[e.g.][]{2013VielWDM,Baur2017,Irsic2017,Irsic2017b,Armengaud2017,Palanque-Delabrouille:2019iyz}, to mention some applications; (iii) study the evolution of the intergalactic medium (IGM) through techniques such as the tomography of the Lyman-$\alpha$ forest~\citep[e.g.][]{Lee:2013gvp,Lee:2014mea,2014Ciweski:Lyatomography,2018ClamatoDR1,Ravoux:2020bpg,Newman:2020iao,Horowitz:2021zwh,Kraljic:2022lzr}. 

For all of these studies, as in general cosmological measurements, it is necessary to characterize possible sources of systematic effects as well as to test analysis pipelines. In this regard, the use of realistic synthetic data sets has acquired an important role over the years. Lyman-$\alpha$ synthetic spectra for BAO analysis and methods to generate them have been used since the analysis of the Baryon Oscillation Spectroscopic Survey (BOSS) year one data~\citep{2011SlosarBOSS} and later on in BOSS Data Release 9 (DR9) \citep{2013BuscaBOSS,2013SlosarDR9}, which used mock data sets following the prescription in \citep{Font2012Mocks}. Later, the analysis of BOSS DR11~\citep{2015DelubacDR11} and DR12~\citep{2017BautistaDR12} used improved mock data developed by \citep{Bautista2015Mocks} which included astrophysical effects such as absorption due to transition lines different from Lyman-$\alpha$. These mocks also used the method of \citep{leGoff2011Sims} to produce a realistic Lyman-$\alpha$ -- QSO cross-correlation and were used to validate the analysis of \citep{2017duMasDR12}. 
The most recent effort on producing Lyman-$\alpha$ mocks was during the analysis presented in extended Baryon Oscillation Spectroscopic Survey (eBOSS) DR16~\citep{2020duMasDR16} including the use of two sets of simulations: \texttt{LyaCoLoRe} and \texttt{Saclay}\footnote{The name makes reference to the institution where these mocks were mainly developed.} mocks. Simulated maps of Lyman-$\alpha$ flux transmission were produced by these two independent methods as described in \citep{Farr2020LyaColore} and \cite{SaclayMocks} respectively. They were both post-processed, using the methodology described in this work, to produce the final synthetic spectra. In all these cases, the role of synthetic quasar spectra has been to test the analysis methodology, to study systematics, and ultimately to validate the BAO measurements.

The currently ongoing Dark Energy Spectroscopic Instrument (DESI)~\cite{DESI:2016fyo,DESI:2016igz} survey is expected to measure the spectra of 40 million galaxies and over a million Lyman-$\alpha$ quasars with redshift $z>2$ in a 14,000~$\rm deg^2$ area during a five-year period, greatly improving the precision of the BAO scale measurements below an accuracy of 1\%~\citep{DESI:2016fyo,DESI:2023dwi}. Achieving this goal requires a robust characterization of systematic errors, necessitating the production of realistic Lyman-$\alpha$ synthetic spectra that accurately capture the statistical properties of the observed data, as well as the effects of astrophysical processes and instrumental noise.

The main goal of this work is to present in detail the methodologies used to construct DESI synthetic spectra datasets for Lyman-$\alpha$ studies, their performance relative to  DESI early data, and the improvements needed in preparation for the DESI year one analysis. In addition, we compare the precision of BAO measurements forecast on mocks with the precision derived from a Fisher matrix analysis. 

This manuscript is organized as follows: \cref{sec:lyaspectra} describes the overall methods used to produce synthetic Lyman-$\alpha$ spectra by using the transmitted flux fraction, quasar unabsorbed spectrum (continuum) template generation and the addition of astrophysical effects. \Cref{sec:surveysimulations} explains the strategies we follow to effectively reproduce a survey in terms of footprint, object density, redshift-magnitude distributions and instrument response model. \Cref{sec:mockvsdata} presents a set of 40 synthetic Lyman-$\alpha$ spectra catalogs that emulate the DESI Early Data Release (EDR)~\citep{DESI:2023ytc} plus 2 months of observations (DESI-M2) and present a qualitative comparison with observed data. \Cref{sec:forecasts} takes advantage of the results of our EDR+M2 mocks to study the use of mocks as a forecast tool for the DESI experiment and predict the ultimate BAO scale constraining power of DESI using mocks of the completed DESI Lyman-$\alpha$ survey sample.

%% file: mockspectra.tex
\section{Simulating \texorpdfstring{Lyman-$\alpha$}{Lya} spectra}\label{sec:lyaspectra}
In this work we will present two kind of synthetic data sets. On the one hand, we present simulations that resemble a particular data release, in particular the DESI EDR+M2 sample, and on the other hand, we present simulations that are designed to reproduce the characteristics of the planned full DESI survey. Although they differ in specifications as the number density of quasars, the region of the sky covered, among other properties, there are several other specifications that are common. Therefore, in this section, we will describe the steps to produce simulated data sets, regardless of which type is to be produced. 

The simulations described here involve, first, the production of flux transmission fields, described in \cref{subsec:rawmock}, and then the addition of astrophysical and instrumental effects with a script named \texttt{quickquasars}\footnote{\url{https://github.com/desihub/desisim/blob/main/py/desisim/scripts/quickquasars.py}}, which is a compilation of DESI code within the \texttt{desisim}\footnote{\url{https://github.com/desihub/desisim}} and \texttt{specsim}\footnote{\url{https://github.com/desihub/specsim}} repositories that are used to simulate realistic spectra. 

Simulating Lyman-$\alpha$ spectra with \texttt{quickquasars} is performed in three main stages. (\textit{i}) The transmitted flux fraction is read from an input mock Lyman-$\alpha$ forest dataset, hereafter referred as raw mocks, e.g. the output of the \texttt{LyaCoLoRe}~\citep{Farr2020LyaColore} or \texttt{Saclay}~\citep{SaclayMocks} programs. Optionally, the transmitted flux fraction can be combined with absorption from Damped Lyman-$\alpha$ Absorbers (DLAs), Broad Absorption Lines (BALs), and additional absorbers, hereafter referred as metals. (\textit{ii}) A template of the quasar unabsorbed spectrum is defined and multiplied by the transmitted flux fraction to create the noiseless quasar template. (\textit{iii}) The instrumental model and observing conditions are introduced to simulate the noisy spectra. The first two of these stages are described in detail in the following sections, while the last will be described in \cref{sec:surveysimulations}. 

\subsection{Transmitted flux fraction and raw mocks}\label{subsec:rawmock}
To simulate the Lyman-$\alpha$ forest, \texttt{quickquasars} uses an input set of transmitted flux fractions, $F$, defined on a set of lines of sight (without instrument noise, continuum template, and astrophysical contaminant features added), which comprise the raw mocks. The transmitted flux fraction is related to the optical depth $\tau$ by $F=e^{-\tau}$. Raw mocks can be produced following different approaches, depending on the scientific use; for modeling the 1D power spectrum $P_{\rm{1D}}$ or the 3D correlations, for instance. In the 1D case the idea is that mocks should reproduce the observed one-dimensional power spectrum~\citep{McDonaldSDSS:2004kjl,Palanque-Delabrouille:2013gaa,Walther:2017cir}. In the 3D case the tracers included in these mocks should be a biased form of the large-scale 3-dimensional matter power spectrum $P_{\rm{3D}}$ usually modeled assuming a Kaiser form~\citep{Kaiser:1987}
\begin{equation}\label{Eq:Pk}
    P_{\rm{3D},i}(\mathbf{k},z)=b_{i}^2(z) (1+\beta_i(z) \mu_k ^2)^2P_L(k,z)
\end{equation}
where $P_L$ is the linear isotropic matter power spectrum, $b(z)$ and $\beta(z)$ are the bias and redshift-space distortion (RSD) parameters, $k^2=k_\parallel^2+k_\perp^2$ and $\mu_k=k_\parallel/k$ where $k_\perp$ and $k_\parallel$ are the components of k perpendicular and parallel to the line of sight. The suffix $i$ denotes the type of tracer modeled, for example: Lyman-$\alpha$ forest, quasars, DLAs or metals. 

Throughout this work, we will refer to three different sets of raw mocks: \texttt{LyaCoLoRe}, \texttt{Saclay} and \texttt{Ohio}, which we briefly describe next. The first two are aimed at 3D correlation studies, and the third is aimed at $P_{\rm{1D}}$ analysis. Regardless of the type of raw mocks used, the production of synthetic spectra using \texttt{quickquasars} is the same as long as the input transmissions are stored according to the format described in \cref{appendix:datamodel}. 
 
\paragraph{\texttt{LyaCoLoRe}:}
These mocks were used for the Lyman-$\alpha$ BAO studies in eBOSS DR16~\citep{2020duMasDR16} and were referred as the \texttt{London} mocks. The methodology to produce this set of raw mocks is fully described in~\citep{Farr2020LyaColore}. It starts with the generation of a correlated Gaussian multivariate overdensity field sampled in a three-dimensional mesh, given an input matter power spectrum, using the Cosmological Lofty Realizations package (\texttt{CoLoRe}\footnote{\url{https://github.com/damonge/CoLoRe}})~\citep{Ramirez-Perez:2021Colore}. The positions of the quasars in the mocks are assigned following an input number density $n(z)$ and bias $b(z)$, and placed in density peaks of the Gaussian random field above a given threshold via Poisson sampling. Density fields along the line of sight from the quasars to the center of the box, namely the skewers, are also extracted at this stage. The results are then post-processed with the \texttt{LyaCoLoRe}\footnote{\url{https://github.com/igmhub/LyaCoLoRe}} code, where small-scale fluctuations are added to the skewers in order to reproduce the variance of the Lyman-$\alpha$ forest in the data. A log-normal transformation is applied to the final Gaussian density field and a Fluctuating Gunn-Peterson approximation (FGPA)~\citep{Croft:1997jf} is used to compute the optical depth on each simulated cell. Then, the effects of RSD are added by computing the radial velocity from the gradient of the Newtonian gravitational potential of the density field. This velocity is then used to shift the position of the optical depth in redshift space accordingly. Lastly, the optical depth $\tau$ is transformed into the transmitted flux by the equation $F=e^{-\tau}$. In addition, the \texttt{LyaCoLoRe} post-processing method includes the generation of transmission due to metals and also provides a catalog with the redshift and column densities of DLAs that are correlated with the density field, as briefly described in \cref{subsec:astrophysical_cont}. 

\paragraph{\texttt{Saclay}:}
These raw mocks are generated following the methodology described in~\citep{SaclayMocks}. They are also based on the generation of Gaussian random fields and the use of fast Fourier transforms. However, there are two main differences with respect to the \texttt{LyaCoLoRe} mocks. First, while \texttt{LyaCoLoRe} uses the same matter density field to model the \lya forest and the quasars, the \texttt{Saclay} mocks generate two different fields using different input power spectra, calibrated in such a way that the resulting log-normal fields have the desired linear power spectrum for both quasars and the \lya forest.\footnote{As pointed out in \citep{Youles:2022pgf}, the auto-correlation function of quasars in \texttt{LyaCoLoRe} mocks is too high on small scales.} Second, in the \texttt{Saclay} mocks the RSD effects are generated by calculating the velocity gradient along each line of sight and then using a modified form of the FGPA that accounts for the gradient. These mocks also include the generation of a catalog of DLAs that are correlated with the density field as in the \texttt{LyaCoLoRe} mocks, but do not include metal absorption, which may be added later by \texttt{quickquasars}.

\paragraph{\texttt{Ohio}:}
These mocks are produced following the procedure described in \cite{Karacayli:2020aad}. They are also based on generating Gaussian random fields, but in this case the skewers are independent and aim to reproduce the expected evolution of the mean transmission from \citep{Faucher-Giguere:2007peq} and a one-dimensional power spectrum comparable to \citep{Walther:2017cir} and \citep{Palanque-Delabrouille:2013gaa}. While \texttt{LyaCoLoRe} and \texttt{Saclay} mocks have quasar positions simulated in such way as to recover the input quasar bias, the \texttt{Ohio} mocks quasars have the same magnitude and are placed in the exact coordinates and redshift as the EDR+M2 observed data, therefore having null value for the Lyman-$\alpha$ -- QSO cross-correlation. The RSD effect is not included, nor is the absorption from DLAs or metals. As mentioned above, these mocks are aimed at P1D studies, which will not be covered in this paper. We refer the reader to \citep{Karacayli:2023afs} and \citep{Ravoux:2023pgi} for further discussion. However, given their flexibility to assign the same redshift, magnitude and exposure time as the EDR+M2 observed data, these were used to assess how well our simulations reproduce the observed signal to noise. 

\subsection{Continuum templates}\label{subsec:continuum}
The quasar continuum template, defined as the unabsorbed spectrum without noise added, is generated by \texttt{quickquasars} using one of the following options:
\begin{itemize}
    \item \textbf{SIMQSO (default):} Templates are generated using the $\texttt{simqso}$ library\footnote{\url{https://github.com/imcgreer/simqso}\label{footnote:simqso}}~\citep{2013:DR9QLF,simqso:2021}, which contains a very broad set of tools to generate mock quasar spectra. The model used has two main components: a broken power-law continuum, and a set of Gaussian emission lines defined by their wavelength, equivalent width, and Gaussian RMS width ($\sigma$). To emulate the power-law continuum component, we randomly generate the slopes of the broken power-law continuum model following a Gaussian distribution centered at the slope value $m$ with dispersion $\sigma_m$ in a particular rest-frame wavelength region.  We use the default values from \texttt{simqso} outside the Lyman-$\alpha$ region which are based in BOSS DR9:  $m=-0.37$ at $5700\ \AA<\lambda<9730\ \AA$, $m=-1.70$ at $9730\ \AA<\lambda< 22300\ \AA$ and -1.03 at $\lambda>22300\ \AA$, all with dispersion $\sigma_m=0.3$. For the Lyman-$\alpha$ forest region we tuned the slopes and dispersion to qualitatively reproduce the mean continuum and the dispersion (obtained from the first five components of a PCA analysis applied quasar spectra~\citep{rodrigo_tesis}) measured in the eBOSS DR16 quasar catalog. The resulting slopes are: $m=-1.5$ at $\lambda<1100\ \AA$ and $m=-0.5$ at $1100\ \AA<\lambda<5700\ \AA$, both with dispersion $\sigma_m=0.7$. To emulate the emission line component, we use the model described in detail in
    \cref{appendix:emissionlines}.
    The model combines emission lines within the Lyman-$\alpha$ forest region from the composite model of BOSS spectra~\citep{Harris:2016ymq} and emission lines outside the Lyman-$\alpha$ region from the \texttt{simqso} model. The emission line diversity is given mainly by the scatter in the equivalent width. Furthermore, the specific values for the mean equivalent width of some of the emission lines were adjusted so that the mean continuum resembles the results from DESI EDR+M2 data. Note that the emission line parameters used in this paper slightly differ from those used to generate the eBOSS DR16 mocks~\citep{2020duMasDR16}. See the left panel of \Cref{fig:deltaattribs} for a qualitative comparison of the mean continuum obtained in our analysis of the simulated spectra versus and the EDR+M2 data, which was also used to make the aforementioned adjustments of the slopes and the emission line parameters. Information about the continuum template generated for each mock quasar, such as the slopes and emission lines equivalent widths, are stored in one of the output "truth" files from \texttt{quickquasars} which are described in \cref{appendix:output}.  
    
    \item \textbf{QSO:} Templates are generated using previously calculated eigenvalues and eigenvectors from a principal component analysis (PCA) decomposition of quasars from SDSS/DR7 ($z=0.4-2$) and BOSS/DR10 quasar spectra ($z=2-4$)~\citep{QSOtemplates_technote}. For each QSO a set of linear combinations of eigenvectors is generated to construct the templates, discarding those with negative flux. The templates are normalized to give the appropriate magnitude, and the sample is reduced to keep only those that could pass the DESI color cuts~\citep{Chaussidon:2022pqg}. Finally, one template is randomly chosen and the corresponding coefficients are stored in the truth files. It is important to mention that this method is not used currently for any use of the mocks, however we briefly describe it because we may implement new templates based on DESI data in the future.  
\end{itemize}

\subsection{Astrophysical contaminants}\label{subsec:astrophysical_cont}
Observed quasar spectra include features due to different astrophysical effects that contaminate the correlation function. In order to study their impact, these features may be added to the generated synthetic spectra. In the following we describe these contaminants.
\subsubsection{Broad Absorption Lines systems}
Quasars with broad ultraviolet (UV) absorption in their spectrum are called Broad Absorption Line (BAL) quasars. This absorption is produced by surrounding gas clouds near the quasar nucleus. BAL quasars are further classified on the basis of spectral lines that show troughs typically blueshifted from their rest-frame wavelength and with velocity widths larger than $2000 \, \rm{km/s}$~\citep{Weymann:1991}. For instance, HiBALs are BALs with absorption in high-ionization lines such as C IV(1549), LoBALs are those with absorption troughs from low-ionization features such as Mg II, whereas FeLoBALs exhibit absorption in lines such as Fe II~\citep{Guo:2019bal,Filbert:2023inprep}. 

BALs are included in the simulated spectra using a set of 1500 templates selected from the BAL catalog produced in \citep{Guo:2019bal}, which in turn classified 53,760 BALs from 320,821 (16.8\%) quasars with a redshift in the range $1.57 < z < 5.56$ from the SDSS DR14 quasar catalog. The BAL templates used for our mocks cover the rest-frame wavelength region of $944.6\ \AA<\lambda < 1686.5\ \AA$ and add BAL features associated with O VI(1031), O VI(1037), Lyman-$\alpha$, N V(1240), Si IV(1398), and C IV(1549).

BAL features are intrinsic to the quasars, and span a wide range of velocity shifts that are not expected to correlate with neighboring quasars. Therefore, in \texttt{quickquasars} we simply select random BAL templates and use them to modify the flux of a certain fraction of randomly selected quasars. Before combining the total flux fraction with the continuum template generated in \cref{subsec:continuum}, it is multiplied by flux fraction of the BAL template. 

The fraction of BAL quasars in the sample is an input parameter, typically set to 16\%, in order to resemble what was found in the DR14 observed data~\citep{Guo:2019bal}. As with other features in \texttt{quickquasars}, important information about BALs is stored in the truth files to recover the templates when needed; for instance: the unique template identification number, as well as various measured properties of the BALs from which the templates were derived, such as Absorption Index (AI), Balnicity Index (BI), etc.

\subsubsection{High Columns Density systems}
The Lyman-$\alpha$ forest primarily arises from neutral hydrogen gas in the intergalactic medium (IGM) along the line of sight. Condensed systems can form additional strong absorption features. Systems with column densities $\log N_{HI} > 17.2\ \text{cm}^{-2}$ are called High Column Density systems (HCDs) and further classified as Lyman Limit Systems (LLS) for column densities within the range $17.2\ \text{cm}^{-2} < \log N_{HI} < 20.3\ \text{cm}^{-2}$ or Damped Lyman-$\alpha$ Systems (DLAs) when the column density exceeds $20.3\ \text{cm}^{-2}$ ~\citep{Wolfe1986}. DLAs produce  features in the Lyman-$\alpha$ forests that can be detected in forests of sufficiently high signal-to-noise, for example using machine learning algorithms~\cite{parks2018deep,ho2020detecting,chabanier2022completed,wang2022deep}. Once detected, they can be masked out of the Lyman-$\alpha$ spectra, as was done in \cite{2020duMasDR16}. However, undetected HCDs still have a significant impact on the measurement and modeling of the Lyman-$\alpha$ correlation function~\citep{Font:DLAs2012}. Thus, having an accurate simulated distribution of HCDs is essential for the generation of realistic mocks.

The default $N_{HI}$ distribution in the mocks is computed following the column density distribution function model from the IGM physics package \texttt{pyigm}\footnote{\url{https://github.com/pyigm/pyigm}}~\citep{prochaska2014towards}. The public repository of the DESI code \texttt{desisim} contains a copy of the tabulated column density distribution function from the \texttt{pyigm} repository. At the mock mean quasar redshift of 2.35, there are approximately 0.39 HCDs and 0.054 DLAs per forest (quasar rest-frame wavelengths between $1040\AA$ and $1200\AA$).

The addition of HCDs to the spectra with \texttt{quickquasars} can be done in two different ways as follows.
\begin{itemize}
  \item \textbf{Random}. The simplest method is to add a random number of HCDs to the transmission at random wavelengths. In this case, the generated HCDs would not be correlated with the density field that was used to produce the raw mocks, which makes this method unsuitable for our purposes and therefore is not used for any of the mocks produced for this work. However, this method is being used in \texttt{Ohio} mocks to study the effect of DLAs in P1D analyses.

  \item \textbf{Correlated}.  
  HCDs are biased tracers of the matter density field, therefore a more realistic approach is to place them in peaks of the density field.
  We follow the method described in detail in section 2.3 of \citep{Font:DLAs2012}, a summary of which we describe here. First, we identify peaks in the Gaussian density fields above a given threshold set by an input bias for HCDs through equation A.12 of Appendix A of \citep{Font:DLAs2012}. Second, we populate these peaks with HCDs following the column density distribution discussed above. Then, we use \texttt{quickquasars} to read the information from the transmission files and inject the HCD absorption into the transmission by using a Voigt profile. Both \texttt{Saclay} and \texttt{LyaCoLoRe} raw mocks assumed a constant bias $b_{\rm{HCD}}(z)=2$~\citep{Perez-Rafols:2017mjf} and assign column densities following the $(N_{\rm{HI}},z)$ distribution of \texttt{pyigm} as in the random DLA method case.
\end{itemize}

\subsubsection{Contamination from additional absorption transitions}
Besides neutral-hydrogen, the IGM contains other absorbers, which we refer to as metals. Because our data analysis protocol transforms wavelengths to redshifts assuming Lyman-$\alpha$ absorption, neutral hydrogen and a metal occupying the same physical position are absorbed at different apparent redshifts, resulting in an apparent distance separation given by
\begin{equation}\label{eq:metals_rparallel}
r_\parallel = (1+z)D_H(z)\Delta \lambda/\lambda_{Ly\alpha},
\end{equation}
where $D_H$ is the Hubble distance and $\Delta \lambda = \lambda_m - \lambda_{Ly\alpha}$ is the wavelength separation of the metal transition ($\lambda_m$) with respect to Lyman-$\alpha$. The auto and cross-correlation functions will then show peaks at $r_\perp\approx 0$ and the $r_\parallel$ scale of each metal of sufficiently strong absorption. The most important such metals are Si II(1260), Si III(1207), Si II(1193), and Si II(1190) corresponding respectively to $r_\parallel=104,\,-21,\,-54$ and $-61 {\ \mathrm{Mpc/h}}$ at $z=2.35$.
 
In addition to this effect due to metal -- neutral-hydrogen correlations, particularly strong absorbers can significantly affect the measured auto-correlation through their own (metal--metal) auto-correlation. The most important transitions are C IV(1548) and C IV(1550). However, these are not included in our mocks because absorption found within the Lyman-$\alpha$ forest region caused by these metals is produced by C IV at redshifts as low as $z\approx1.32$, outside the range of the transmission files of the raw mocks.

Given its impact in Lyman-$\alpha$ forest BAO analysis it is important to characterize their contribution to the correlation functions. To perform this task we can include metals in the simulated spectra using one of the following methods: 

\begin{itemize}
\item \textbf{Added by \texttt{quickquasars}.} 
In this case, we take the Lyman-$\alpha$ transmitted flux fraction at rest-frame as it is given in the transmission files. The optical depth of the Lyman-$\alpha$ absorption is then re-scaled by a factor $C_m$ according to the relative absorption strength of each metal with respect to Lyman-$\alpha$. The transmitted flux fraction of the metals is then computed as $F_{m} = e^{-C_m \tau_{Ly \alpha}}$. Finally, the absorption features are moved to their corresponding observed wavelengths according to the relation $\lambda_{obs,m} = (1+z_{abs})\lambda_m$, where $z_{abs}=\lambda/\lambda_{Ly\alpha}-1$ is the redshift at which the absorption occurs. In this method, the re-scaling of the optical depth is done after RSD effects have been applied to the Lyman-$\alpha$ absorption forcing the metals to have the same RSD parameter as Lyman-$\alpha$.
The absorption amplitude of each metal transition strongly affects the value for its corresponding bias $b_m$ ($b_i$ defined in \cref{Eq:Pk}, for metals). In other words, the bias highly depends on the intensity of the optical depth of each independent transition and consequently on the value $C_m$ used in this method. 

For this simple simulation technique, $b_m$ and $C_m$ have a linear relation. This allows to find the $C_m$ needed to reproduce the measured biases of different eBOSS datasets. For this purpose, we made several realizations of the same mock including Lyman-$\alpha$ absorption, and absorption contamination due to Si~II(1260), Si~III(1207), Si~II(1193), and Si~II(1190). We varied the value of $C_m$ for each realization and measured the corresponding metal biases $b_{\eta,m}$. Then, we found their linear relation through a simple linear regression, and interpolated the $C_m$ value needed to obtain a target bias. In this case, we chose the absorption coefficient value that reproduced the results of eBOSS DR14~\citep{2019deSainteAgatheDR14}. We performed this procedure on the Lyman-$\alpha$ transmission files of the \texttt{LyaCoLoRe} and \texttt{Saclay} mocks independently, the resulting $C_m$ values are shown in \cref{tab:bias_coef}. We set these values as default in \texttt{quickquasars} when using this method. However, other values of the $C_m$ parameters can be provided as an input to \texttt{quickquasars} if required. Note that the values for \texttt{LyaCoLoRe} are different that those of \texttt{Saclay} mocks in \Cref{tab:bias_coef}, this is expected since the values of the $C_m$ coefficients depend on the input details to construct the raw mocks, such as the flux-transmission distribution and the cell size used, to mention some.

The $C_m$ values results were validated by computing the bias for a sample of ten \texttt{LyaCoLoRe} and ten \texttt{Saclay} mock realizations and were found to be statistically consistent with the target DR14 bias value of each metal. We refer the reader to \citep{andrea_tesis} and \citep{andrea_proceedings} for further details about these results. The same values were also used to produce the mocks presented in the eBOSS DR16 analysis~\citep{2020duMasDR16}, also presenting consistency between mocks and observed data. 

\begin{table}[!tbp]
\centering
\caption{Coefficient $C_m$ values of Si~III(1207), Si~II(1190), Si~II(1193), and Si~II(1260) used by default when including metals through the \texttt{quickquasars} method for \texttt{LyaCoLoRe} and \texttt{Saclay} mocks.}
\label{tab:bias_coef}
{\footnotesize
\begin{tabular}{p{1.7cm}ll}
   &  \multicolumn{1}{c}{\texttt{LyaCoLoRe}}  & \multicolumn{1}{c}{\texttt{Saclay}} \\
 \multicolumn{1}{c}{\textbf{Transition}} & \multicolumn{1}{c}{$\mathbf{C_m}$} & \multicolumn{1}{c}{$\mathbf{C_m}$}  \\
\hline
\hline 
 Si II(1190) & $6.4239\times 10^{-4}$ & $4.4960\times 10^{-4}$  \\ 
 Si II(1193) & $9.0776\times 10^{-4}$ & $6.3540\times 10^{-4}$  \\ 
 Si II(1260) & $3.5420\times 10^{-4}$ & $4.2504\times 10^{-4}$  \\ 
 Si III(1207)& $1.8919\times 10^{-3}$ & $9.4595\times 10^{-4}$  \\ 
\end{tabular}
}
\end{table}

 \item \textbf{From the transmission files.} In this method the transmitted flux fraction of metals is computed during the generation of the Lyman-$\alpha$ transmission in the raw mocks. An example of this are the \texttt{LyaCoLoRe} raw mocks that included metal transmissions on their data products using the methodology explained in section 5.2 of \citep{Farr2020LyaColore}, which we briefly describe here. Similarly to the method followed in quickquasars, here it is assumed that the optical depth of each additional metal is proportional to that of the Lyman-$\alpha$ absorption. Therefore for each metal we need to define a relative absorption strength $A_m$, so that $\tau_{m}=A_{m}\tau_{Ly \alpha}$ and a rest-frame wavelength at which the metal absorption will be included. Notice that in general, the absorption strength $A_m$ here is not the same as the coefficient $C_m$ mentioned above. However, the \texttt{LyaCoLoRe} mocks produced in \citep{Farr2020LyaColore} made use of the coefficients found using the \texttt{quickquasars} method, described above.\footnote{All the relative absorption strength values used to produce the metal transmissions of the \texttt{LyaCoLoRe} mocks correspond to the $C_m$ coefficients found with the \texttt{quickquasars} method except for Si~II(1190) where a value $A_{\text{Si~II(1190)}}=1.28478\times 10^{-4}$ was used to reach the target bias value of eBOSS DR12 $b_{\text{Si~II(1190)}}=-4.4\times 10^{-3}$~\citep{2017BautistaDR12} (see Tables 2 and 3 in \citep{Farr2020LyaColore}).} Finally, RSDs are applied and the transmitted flux fraction of each metal is computed separately. The aforementioned Si and Lyman-$\beta$ transmissions are then saved independently in the transmission files, which can then be read by \texttt{quickquasars} in the same way as the Lyman-$\alpha$ transmissions and added to the total transmitted flux. 
 \end{itemize}

\section{Emulating a survey }\label{sec:surveysimulations}
\subsection{Survey demography}\label{subsec:survey_attribs}
In general terms, the generation of synthetic spectra, as presented in previous work \citep{2011SlosarBOSS,2013BuscaBOSS,2013SlosarDR9,Font2012Mocks,Bautista2015Mocks,2020duMasDR16} was done for a specific region of the sky, for instance, the whole eBOSS or DESI footprints, with a uniform density of QSOs and the same exposure time for all of them. This is appropriate to mimic the final stages of the given experiment, for which the whole of its footprint has been scanned according to its design plans, but a different approach is necessary for ongoing experiments in their early stages of observation.
To address this, \texttt{quickquasars} can use the following as inputs.

\begin{figure}
    \centering
    \includegraphics[width=\textwidth]{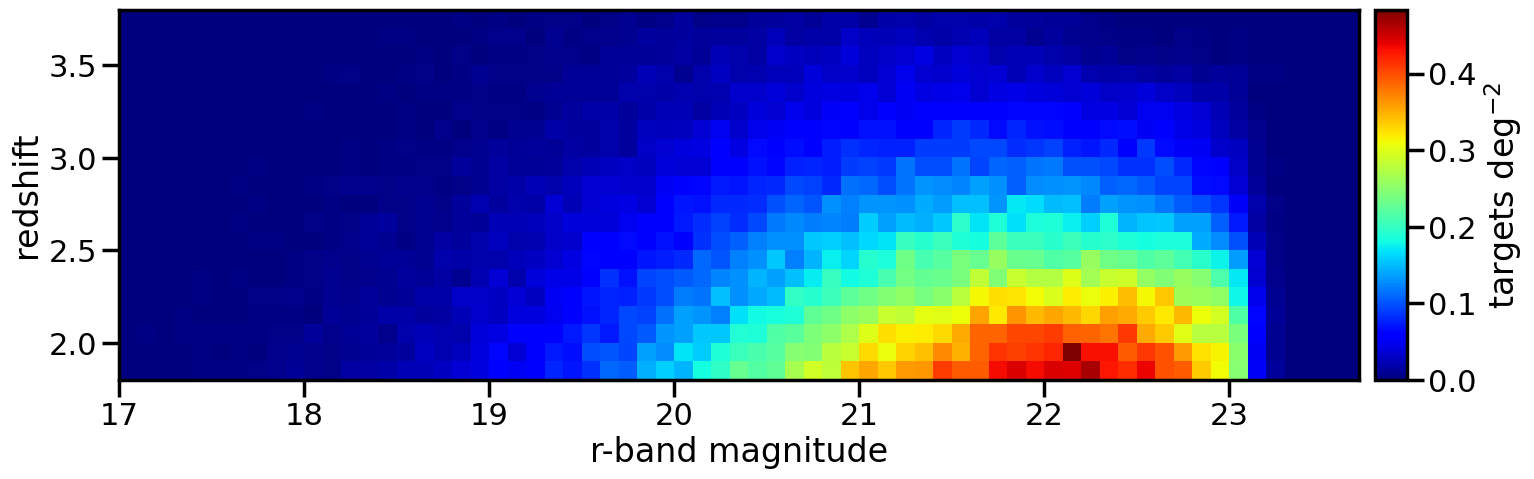}
    \caption{Quasar number density distribution as a function of redshift and magnitude as expected by the DESI quasar target selection pipeline\citep{Chaussidon:2022pqg}. The color scale gives the number of quasars per $\rm deg^2$
    and per redshift-magnitude bin ($0.1\times0.1$) with a total
    of $\approx100{\rm deg^{-2}}$. This distribution is used as an input to \texttt{quickquasars} to produce DESI mocks.}
    \label{fig:EDRQLF}
\end{figure}

\paragraph{Redshift-Magnitude distribution.} The default method implemented in \texttt{quickquasars} is to read the quasar redshifts as were assigned during the raw mock generation procedure, and then assign a random magnitude during the \texttt{simqso} continuum template generation method (described in \cref{subsec:continuum}) by following the Quasar Luminosity Function (QLF) from BOSS DR9~\citep{2013:DR9QLF}. While this QLF could be updated to use the results from more recent studies, e.g \citep{Palanque:2016QLF}, here we propose and implement a simpler approach. 
First, we compute a tabulated quasar number density distribution as a function of redshift and r-band magnitude using the DESI-M2 sample, see \Cref{fig:EDRQLF}. Note that the magnitude and redshift distribution of DESI-M2 are the result of the quasar selection procedure described in \citep{Chaussidon:2022pqg}, and such distributions were found to be consistent with the QLF measured by \citep{Palanque:2016QLF}. Next, since the raw mocks contain a larger quasar number density than the expected for DESI, we randomly down-sample the number of quasars by redshift bin following the aforementioned number density distribution marginalized over magnitudes. Finally, we assign a random r-band magnitude to each quasar according to its redshift and a probability computed from the number density distribution in \Cref{fig:EDRQLF}.

\paragraph{Footprint and quasar density:} By default we simulate over the whole DESI footprint, however if a specific area is to be simulated, such as fractions of the DESI footprint or the footprint from other surveys, we use an input number density function tabulated into HEALpix pixels~\citep{2005healpix} including only those pixels that cover the footprint to be simulated. \texttt{Quickquasars} then uses this information and random samples the number of objects from the available skewers to match the observed number density. We used this sub-sampling method as a first approximation for EDR+M2 mocks only, even though in principle it could affect the Lyman-$\alpha$ -- QSO cross and QSO auto-correlations at large scales for the HEALpix pixel size used.\footnote{We use HEALpix pixel of \path{nside=16} which corresponds to a $\sim250$ Mpc/h scale at redshift z=2.3.} This sub-sampling method differs from the one used for eBOSS mocks in the DR16 analysis validation~\citep{2020duMasDR16} and therefore does not affect its results. A more suitable method will be studied for future releases of DESI mocks. 

\paragraph{Exposure time distribution:} We include two options to assign exposure times. If it is not expected that all the simulated quasars have the same exposure time, we use an exposure time probability distribution function computed for each HEALpix pixel of the footprint to randomly assign a multiple of 1000 seconds to the selected quasars. Otherwise, we assign all the quasars the same exposure time, 4000 seconds in the case of complete DESI survey mocks.

\subsection{Simulating the spectra}
\label{sec:specsim}
As mentioned above, \texttt{quickquasars} is an assembly of several pieces of DESI code that include actual simulations of the instrument response and observation conditions. In this section, we describe its most important components that allow simulated spectra to be close to those observed by DESI in terms of instrumental noise. Most of the calculations are performed by \texttt{specsim}\footnote{\url{https://github.com/desihub/specsim}}, a Python package for efficient and flexible simulations of the response of a multi-fiber spectrograph.\footnote{A description of this package, and how to use it as stand alone can be found in \url{https://specsim.readthedocs.io/en/latest/}} The \texttt{specsim} package models the effects of the atmosphere and instrument, to convert an input spectral energy distribution (SED) and source profile into arrays of expected mean detected fluxes with associated variances for each arm of the spectrograph. Our synthetic spectra use the noiseless templates for the input SED and assume a point source. \texttt{Specsim} can be configured for different instruments and conditions and currently supports both DESI and eBOSS simulations.

\begin{figure}[!tbp]
  \centering
  \includegraphics[width=0.87\textwidth]{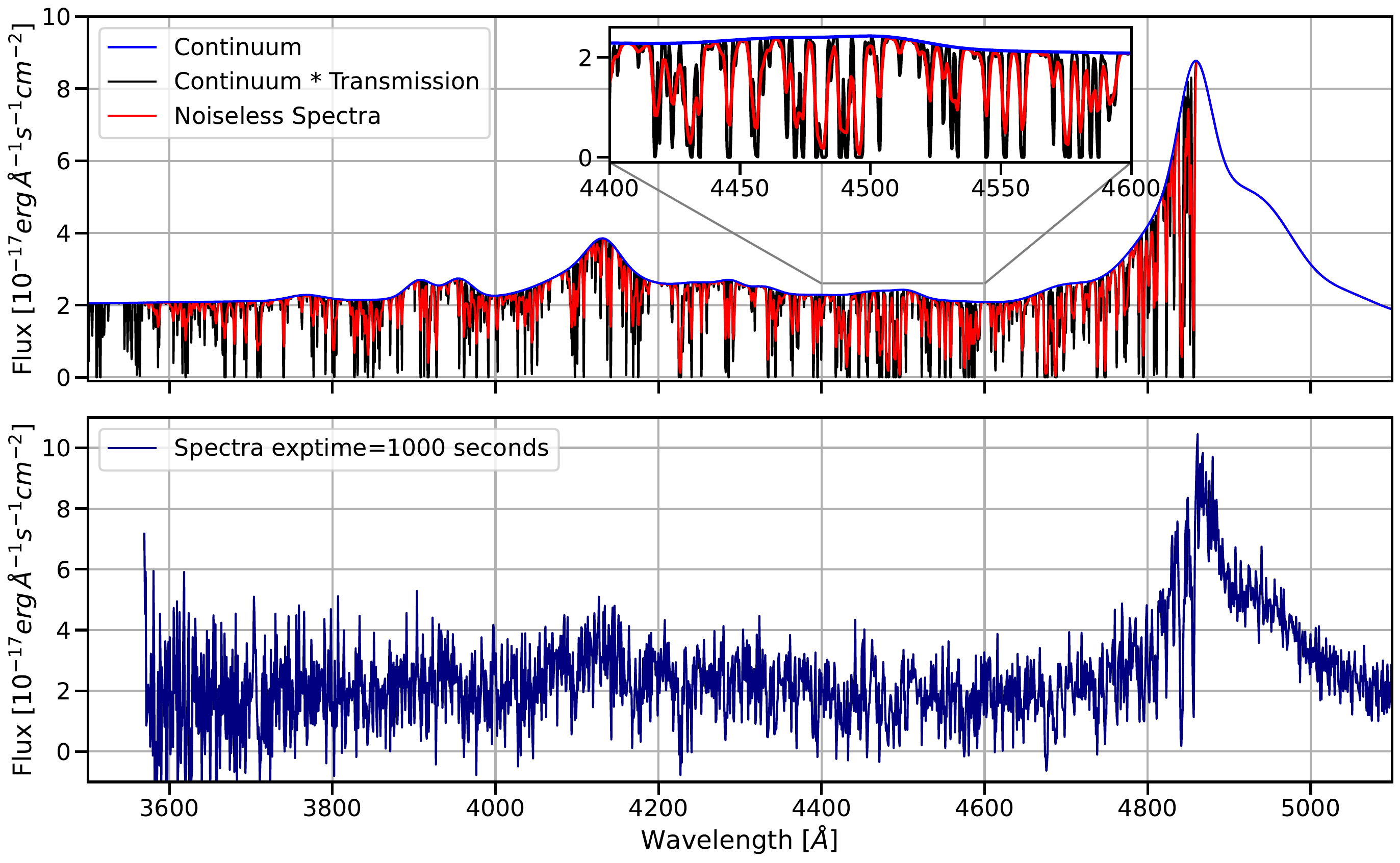}
   \caption{Quasar spectrum through the different stages in \texttt{quickquasars}. Top panel: the blue line shows the continuum template of the quasar before adding Lyman-$\alpha$ absorption, the black line shows the results of multiplying the continuum by the transmitted flux fraction from the raw mocks and red line shows the result of adding the spectrograph resolution to the spectra without instrumental noise added. We also show a zoom in of the $4400 < \lambda < 4600~\AA$ region for better appreciation of the difference between the spectrum lines. Bottom panel: Same spectrum as in the top panel but after instrumental noise has been added emulating 1000 seconds of exposure time.}
  \label{fig:spectra_mock}
\end{figure}

\Cref{fig:spectra_mock} shows an example of a simulated quasar spectrum through the different stages in its production explained in \cref{subsec:rawmock,subsec:continuum}. The bottom panel of this figure includes the same spectrum after passing through \texttt{quickquasars} and adding instrumental noise as will be explained throughout this section.

\subsubsection{The atmosphere}\label{subsubsec:atmosphere}
The atmosphere is modeled by applying extinction and adding sky background, both of which are wavelength dependent. The extinction is scaled with the observing airmass and the sky background includes the effects of the moon when it is above the horizon. The source profile (a delta function representing a point-like quasar in our case) is convolved with an atmospheric point-spread function (PSF) to determine the profile of light incident on the telescope's primary mirror. We use a Moffat profile for the PSF,
\begin{equation}
  I(r)=I_0 \left(1+\frac{r}{\alpha}\right)^{\beta} \, ,
\end{equation}
where $\beta=3.5$~\citep{2020Mayal}, $\alpha = \mathrm{FWHM}(\lambda)/(2\sqrt{2^{1/\beta}-1})$ and $r$ is the on-sky angular distance from the PSF centroid. Here, FWHM stands for the full width half maximum: the radius at which $I(r) = 0.5 I_0$, assumed to scale with wavelength as $\rm{FWHM}(\lambda) = \rm{FWHM_{ref}}*(\lambda / \lambda_{\rm{ref}})^{-1/5}$, where we have set $\rm{FWHM_{ref}} = 1.1 \ \rm{arcsec}$ and $\lambda_{\rm{ref}}=6355\ \AA$.

 \subsubsection{The instrument}\label{subsubsec:instrument}

 \begin{figure}[!tbp]
 \centering
\includegraphics[width=\textwidth]{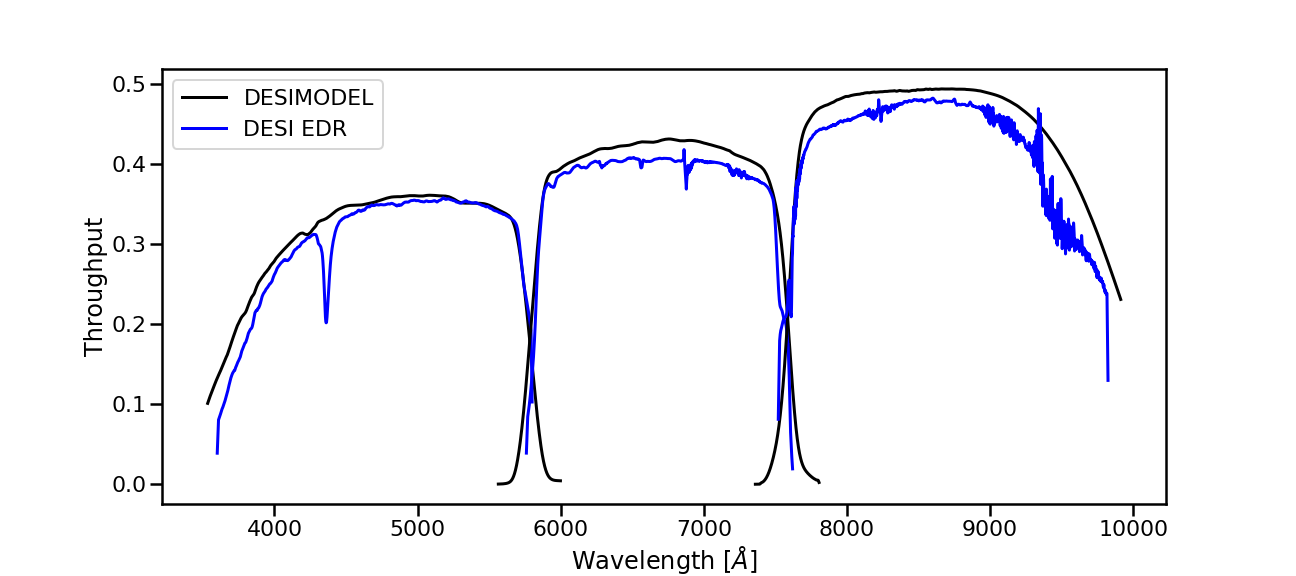}
\caption{Throughput model of DESI spectrograph as expected by the nominal design (black) and as observed during EDR (blue). In the EDR case, there is a dip feature at $4400\AA$ due to an imperfection in the spectrograph blue channel collimator coating~\citep{DESI:2022xcl}. This dip feature is being fixed as the survey progresses and should be present only in spectra collected during the first year of operations.}\label{fig:throughput}
\end{figure}

The incident source flux is multiplied by the effective area of the primary mirror and exposure time. The fraction of light entering the 107$\mu$m-diameter DESI fibers is then estimated using the convolution of the source profile, atmospheric PSF and a wavelength- and field-radius dependent model of the corrector optics PSF. The resulting profile is truncated at the fiber radius, allowing for the expected centering errors due to imperfect fiber positioning and guiding. The sky surface brightness is transformed from sky coordinates to focal-plane coordinates using the spatially varying and anisotropic plate scale to determine the background flux superimposed on the signal. Next, the overall wavelength-dependent system throughput is applied, as shown in figure~\ref{fig:throughput}, which accounts for losses in the primary mirror, corrector optics, fiber transmission, and Charge-Coupled Device (CCD) quantum efficiencies. Finally, wavelength dispersion in each camera is modeled with a sparse convolution matrix. Results are provided as photon fluxes incident on the fibers and as electron and Analog-to-Digital Units (ADU) counts detected by the spectrograph cameras.

\texttt{Specsim} calculates a variance for each output wavelength bin that accounts for the shot noise from the source and sky background, thermal dark currents in the CCDs and noise injected by the readout electronics. Dark currents and read noise account for the effective number of CCD pixels per wavelength bin due to the dispersion of light. The assumed DESI readout noises by camera are 3.29 (b), 3.69 (r), 3.69 (z)  electrons per CCD pixel, but have negligible impact on the relatively long dark-time exposures of QSO targets. The assumed DESI dark currents are negligible at 1.89 (b), 1.14 (r), 1.14 (z) electrons per hour per CCD pixel. 

\subsubsection{Variance smoothing}
The estimation of the pixel variance of the CCD active region in the DESI spectroscopic data processing pipeline~\citep{Guy:2022wlv} includes the contribution of the readout noise variance and a Poisson variance. This method couples the CCD pixel value to its estimated variance, which leads to biases in the resulting calibrated spectrum of the targets. The DESI spectroscopic pipeline addresses this problem by smoothing the sky subtracted spectrum of each target with a convolution using a Gaussian kernel of $\sigma=10\ \AA$ with outlier rejection.

This smoothing step is not part of \texttt{specsim}, so we emulate this process in \texttt{quickquasars} as follows: We subtract the source electron shot noise contribution from the total variance estimated by \texttt{specsim}, explained in \cref{subsubsec:instrument}. We smooth this source electron contribution with a Gaussian kernel using Fast Fourier Transforms (FFTs), where the input array is padded with boundary values to prevent periodicity from distorting values at the edges. Finally, this smoothed source contribution is added to the total variance. The standard deviation of the Gaussian kernel $\sigma$ is an input parameter in \texttt{quickquasars}, which we set to $10~\AA$ by default based on the DESI spectroscopic data processing pipeline.

 \subsubsection{Observing conditions}\label{subsubsec:obsconds}
 The atmosphere model presented in \cref{subsubsec:atmosphere} depends on the assumed observing conditions such as seeing, air mass, moon illumination fraction, moon altitude and moon separation from the tile being simulated. For Lyman-$\alpha$ mocks we use the DESI dark-program optimal conditions setup where the seeing is set to $1.1\ \rm{arcsec}$, air mass to 1.0, and we set the moon illumination fraction to 0 while the moon altitude and separation from the tile are set to $-60\ \rm{degrees}$ and $180\ \rm{degrees}$, respectively. While these conditions might not be particularly realistic, they correspond to how the effective exposure time is defined for DESI observations. In other words, a 1000 seconds exposure simulated at these conditions provides the expected signal-to-noise ratio that an actual observation under different conditions aims to achieve.

\subsubsection{The resolution matrix} 
The DESI spectrograph resolution is encoded in our data products in the form of a "resolution matrix" that is saved in the output spectra files. In the observed data case, this matrix is a result of the spectroscopic extraction algorithm and encodes the dispersion along wavelength due to the spectrograph finite PSF and the CCD pixel size. More details can be found in section 4.5 of \cite{Guy:2022wlv}. This band-diagonal matrix \footnote{The resolution matrix elements far from the diagonal are zero since it combines only flux values from neighboring wavelength.} applied to a high resolution spectrum results in a spectrum at the resolution of the observations. For mocks, each row of the resolution matrix can be approximated as a Gaussian of parameter $\sigma = (\sigma_{LSF}^2 + \Delta \lambda^2/12)^{1/2}$, where $\sigma_{LSF}=0.73$\AA\ at 4000\AA\ and $\Delta \lambda=0.8$\AA\ is the output wavelength bin size. 

In the context of the fast simulations performed here, we adapt the sparse resolution matrix used by \texttt{specsim} to the format of our data products. If required, we store in the truth files the matrix only once per HEALpix pixel file instead of saving for each fiber as is done with real data, because the same resolution is applied to all fibers in the simulations.

%% file: earlymocks.tex
\section{DESI EDR+M2  \texorpdfstring{Lyman-$\alpha$}{Lya}  mocks comparison with data}\label{sec:mockvsdata}
In this section, we present the DESI Lyman-$\alpha$ EDR+M2 mocks, a collection of 40 different mock datasets: 20 for \texttt{LyaCoLoRe} and 20 for \texttt{Saclay} raw mocks, of which 10 mocks contain only Lyman-$\alpha$ absorption and 10 additionally include HCDs, 
BALs and metals. The metals were extracted directly from the transmission files for \texttt{LyaCoLoRe}, and added by \texttt{quickquasars} for \texttt{Saclay}. We include absorption due to Lyman-$\beta$, Si~II(1190), Si~II(1193), Si~II(1260) and Si~III(1207). 

To produce all of these mocks, we extracted the observed footprint, quasar number density, and exposure time distribution from quasar catalog used for the DESI EDR+M2 Lyman-$\alpha$ 3D correlation functions measurement~\citep{Gordon:2023inprep}. This catalog combines the data of DESI Early Data Release (EDR), collected during the survey validation phase, and the first two months of the Main Survey (DESI-M2). The resulting observed data sample consists of 318k QSO targets within the $0<z<6$ redshift range. Next, we show a comparison of the results of these mocks to those of the observed data. 

\subsection{Demographics}\label{subsec:demographics}
\begin{figure}[!tbp]
    \centering
    \includegraphics[width=\textwidth]{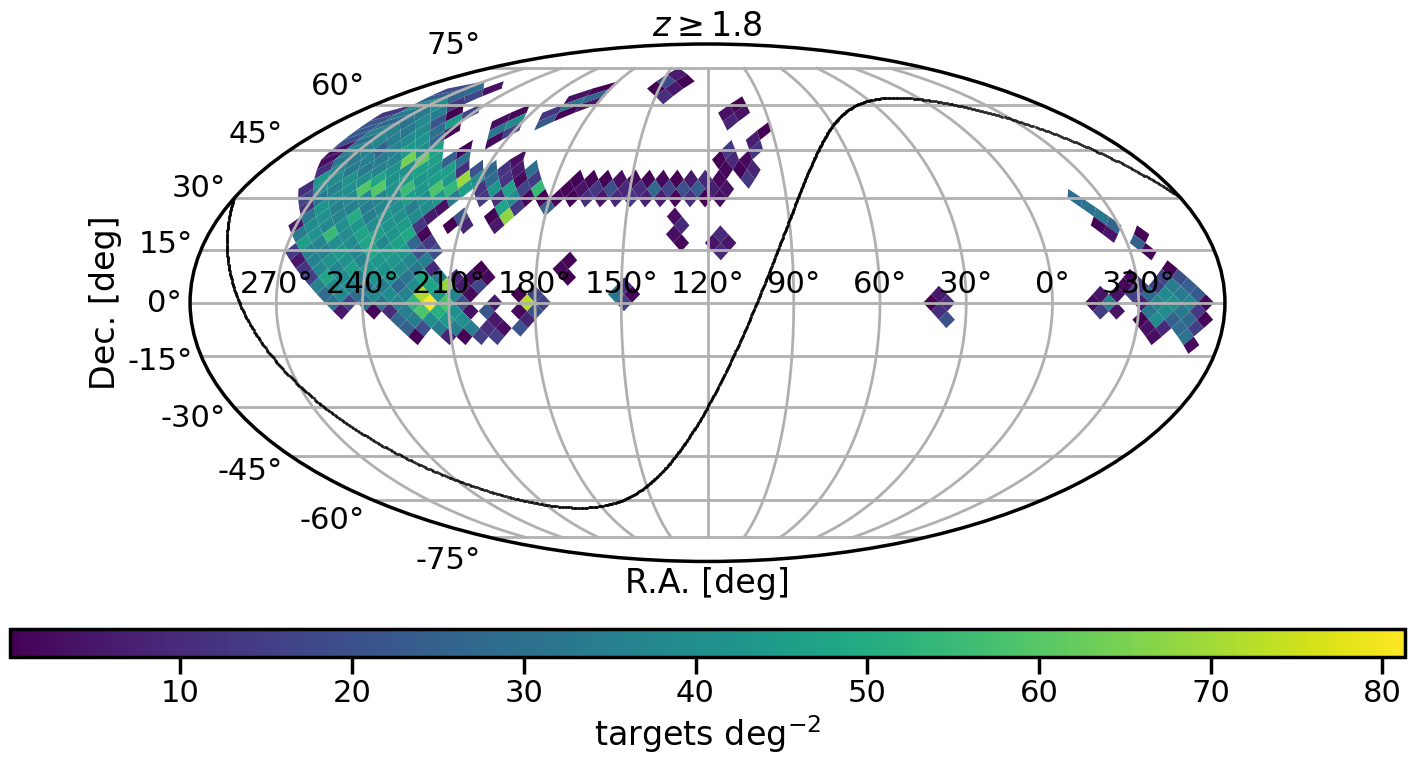}
    \caption{DESI EDR+M2 observed data footprint in HEALpix pixel representation of \texttt{nside=16}. Color shows the $z\geq 1.8$ quasar density. By construction our mocks follow the same footprint and density.}
    \label{fig:density}
\end{figure}

For mocks we only consider QSOs with redshift $1.8<z<3.8$, which results in a sample of 141k targets including 107k Lyman-$\alpha$ QSOs at $z>2.1$. We ignore targets in the regions explored during the EDR phase that fall outside the main DESI footprint (far north and far south) since these regions are not considered by the raw mocks. This makes no impact on our comparison since these targets represent only 0.9\% of the sample. The resulting footprint of the EDR+M2 dataset with the considerations previously mentioned is shown in \cref{fig:density}. By construction all of the produced mocks follow the same footprint, QSO number density and redshift-magnitude distributions as the observed data with negligible variations between realizations.

\begin{figure}[!tbp]
    \centering
    \includegraphics[width=0.49\textwidth]{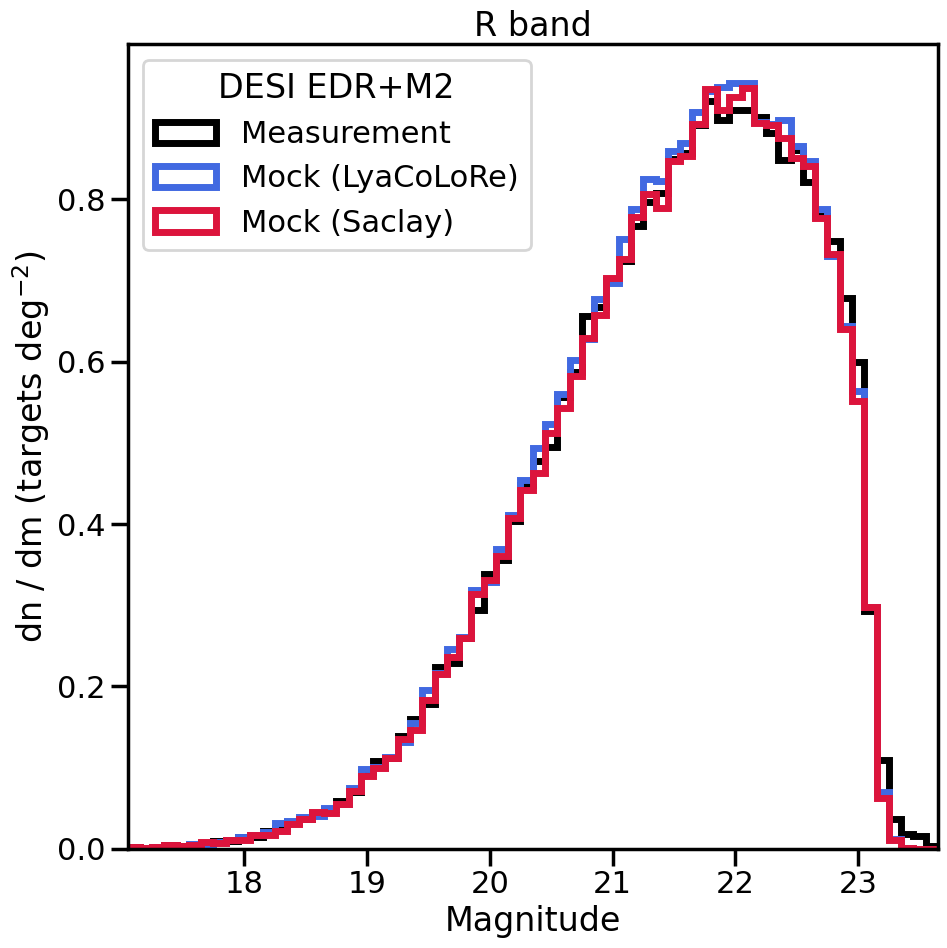}
    \includegraphics[width=0.5\textwidth]{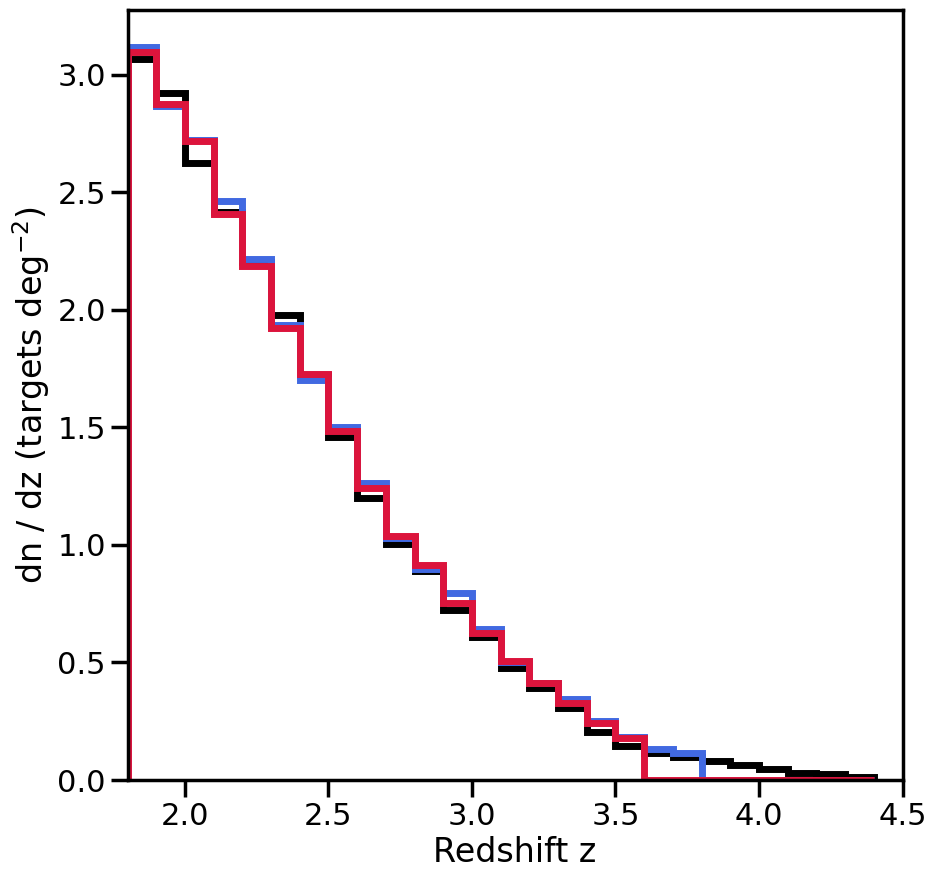}
    \caption{Observed r-band magnitude (left) and redshift (right) distributions of the EDR+M2 data along with the results obtained by one realization of the \texttt{LyaCoLoRe} (blue) and one of the \texttt{Saclay} (red) mocks.}
    \label{fig:nmag_nzr}
\end{figure}

As mentioned in \cref{subsec:survey_attribs}, our mocks follow by construction the redshift-magnitude distribution from \cref{fig:EDRQLF} which is slightly different from that measured on the EDR+M2 sample. This is due the fact that the EDR sample includes fainter targets than those expected to be measured during the main DESI survey (e.g DESI-M2).\footnote{During the survey validation phase an extension of the r-band magnitude limit was tested to study the redshift distribution and population of fainter objects.} However, this does not affect the quality of our mocks since the amount of this type of target represents a small fraction of the sample as can be seen in the left panel of \cref{fig:nmag_nzr} at $M_r>23.1$. 

\Cref{fig:nmag_nzr} shows the observed EDR+M2 data redshift (right) and r-band magnitude (left) distributions, and the corresponding results of one mock realization of each \texttt{LyaCoLoRe} and \texttt{Saclay} mocks. The differences between mocks and data distributions are below the 1\% level and we obtain negligible dispersion between mock realizations. This validates the appropriateness of our methods of redshift sampling and random r-band magnitude assignation, described in~\cref{subsec:survey_attribs}. We see a cutoff in the redshift distribution of the mocks at $z>3.8$ for \texttt{LyaCoLoRe} and $z>3.6$ for \texttt{Saclay}. This comes from the fact that we reach the redshift limit of the raw mocks at this point. The effect of these limits in our analysis and validations are negligible, since the QSOs at these redshifts represent roughly 1\% of the observed sample.

To emulate the signal to noise ratio (SNR) of observed data, we assign random exposure times to the simulated quasars by using am exposure time distribution function, as explained in \cref{subsec:survey_attribs}. We obtained the distribution function used for our EDR+M2 mocks from the effective exposure time of observed data defined by spectroscopic data processing pipeline as  $T_{\rm{eff}} = 12.15\ \rm{seconds} \times \text{TSNR}^2_{\rm{LRG}}$, where $\text{TSNR}_{\rm{LRG}}$ is the LRG template signal to noise ratio (see section 4.14 in \citep{Guy:2022wlv}, for further details). Additionally, we use the measured throughput model achieved during the EDR phase (blue line in \cref{fig:throughput}), which includes a dip feature at $\lambda \sim 440$ nm that affects the noise of the simulated spectra~\citep{DESI:2022xcl}. We also consider the effect of galactic extinction on spectra by adding an O'Donnell extinction model~\citep{ODonnell1994ApJ}. This in principle modifies the magnitude of the mock QSOs from the randomly assigned ones, and therefore modifies the resulting magnitude distribution. We address this issue by re-scaling all fluxes by a factor $F=f_0/f_{\rm{EBV}}$, where $f_0$ is the flux randomly assigned by our method, and $f_{\rm{EBV}}$ is the flux after galactic extinction has been applied. 

\begin{figure}[!tbp]
    \centering
    \includegraphics[width=\textwidth]{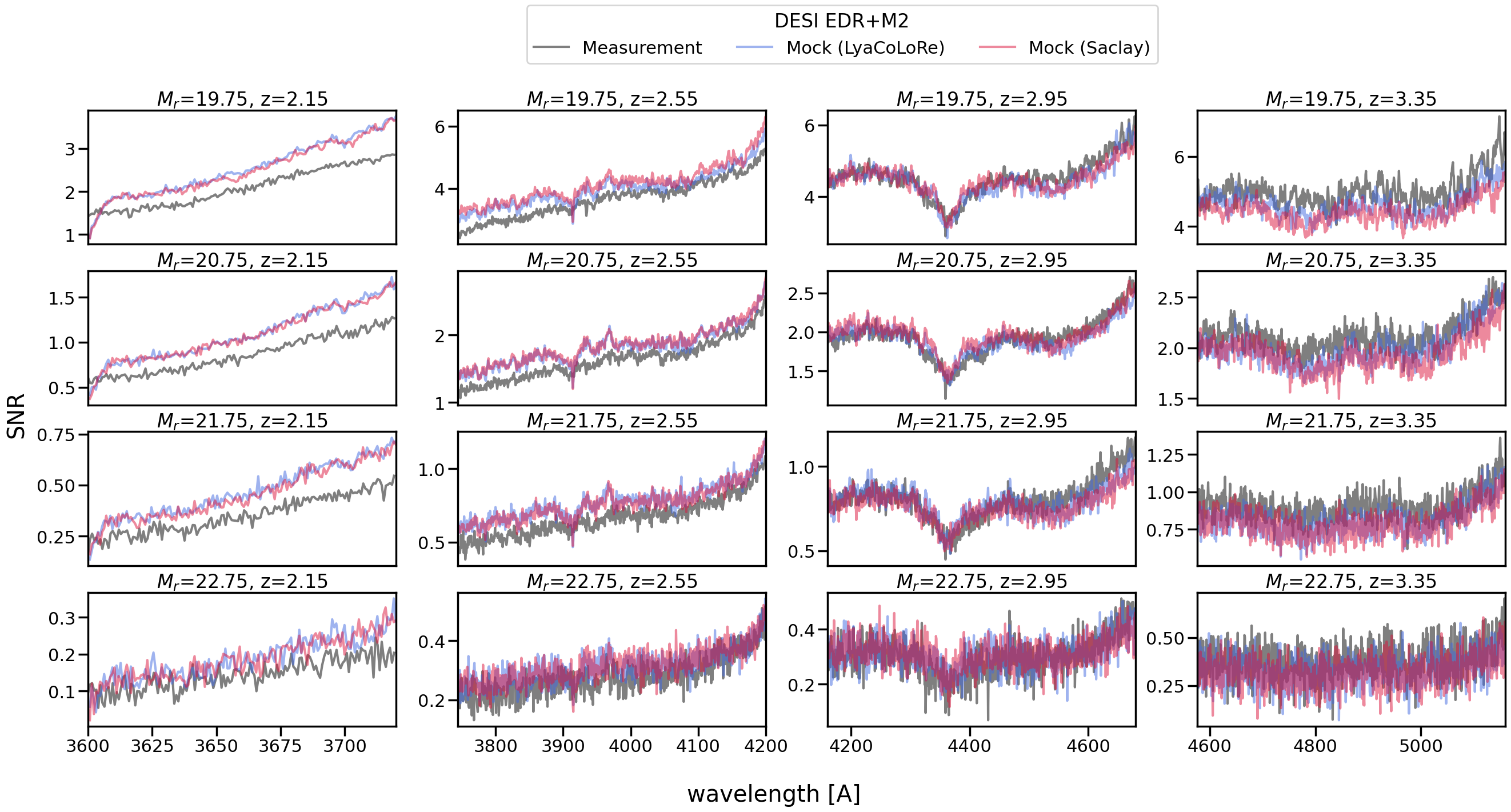}
    \caption{Mean signal to noise ratio (SNR) in the Lyman-$\alpha$ forest region of a 75k QSO sample divided into redshift and r-band magnitude bins. The numbers above each panel represent the central values of the r-band magnitude and redshift bins of widths 0.5 and 0.1, respectively.}
    \label{fig:SNR}
\end{figure}

\Cref{fig:SNR} shows the median SNR of a 75k randomly selected QSO sample in the Lyman-$\alpha$ forest region of observed data and a mock realization of both \texttt{LyaCoLoRe} and \texttt{Saclay} mocks. We have considered r-band magnitude and redshift bins of 0.5 and 0.1 width, respectively. We see a significant difference in the SNR of our mocks compared to data at bright low redshifts quasars ($z<2.2$, $M_r<22$), where we obtain higher SNR in our mocks than in observed data, which might be due to an underestimation in our instrumental noise model for bright low redshift quasars. We expect to study and improve on this issue for future mocks.

\subsection{Correlation functions}\label{subsec:corrfuncs_EDR}
We compute the Lyman-$\alpha$ auto and Lyman-$\alpha$ -- QSO cross correlations following the same procedure as the main analysis~\citep{Gordon:2023inprep} which follows the pipeline analysis used in eBOSS DR16~\citep{2020duMasDR16} with a few modifications in the variance and weighting scheme calculations (see \citep{Ramirez-Perez:2023blu} for further details) and a slightly different wavelength range ($3600\ \AA<\lambda< 5772\ \AA$). This analysis pipeline also masks DLAs and BALs systems found in data. We use the publicly-available package \texttt{picca}\footnote{\url{ https://github.com/igmhub/picca/}}\citep{2021PICCA} to perform this analysis.

First, the flux-transmission field $\delta_q(\lambda)$ is computed from the ratio of the observed flux $f_q(\lambda)$ and the mean expected flux $C_q(\lambda)\bar{F}(\lambda)$,
\begin{equation}
    \delta_q(\lambda) = \frac{f_q(\lambda)}{C_q(\lambda)\bar{F}(\lambda)}-1,
\end{equation}
where $C_q(\lambda)$ is the continuum of the quasar and $\bar{F}(\lambda)$ is the mean transmission. The mean expected flux of each quasar is fitted while measuring the delta field by using the linear approximation
\begin{equation}
     C_q(\lambda)\bar{F}(\lambda) = \bar{C}(\lambda_{\rm{RF}})\left(a_q+b_q\frac{\Lambda-\Lambda_{\rm{min}}}{\Lambda_{\rm{max}}-\Lambda_{\rm{min}}}\right),
\end{equation}
where $\bar{C}(\lambda_{\rm{RF}})$ is the mean rest-frame continuum, $a_q$ and $b_q$ account for quasar spectral diversity, and $\Lambda = \log \lambda$. We have chosen $\lambda_{\rm{min}}=1040\ \AA$ and $\lambda_{\rm{max}}=1205\ \AA$ as in the observed data analysis~\citep{Ramirez-Perez:2023blu,Gordon:2023inprep}. At this same stage some quasar spectra are rejected if they do not meet the quality standards set by the pipeline for reasons such as the forest being too short to be analyzed, failed continuum  fitting, low SNR, to mention some. This results in 88.5k forests in observed data as reported by \citep{Gordon:2023inprep}, while for our mocks we obtain approximately 89.7k analyzed forests for each of the \texttt{LyaCoLoRe} mocks and for \texttt{Saclay} mocks. This 0.7\% relative difference on the number obtained in our mocks with respect to data might be due to the differences in SNR of the mocks compared to data discussed in \cref{subsec:demographics} and also to the shorter redshift range in the case of \texttt{Saclay} mocks.

The variance $\sigma_q$ for the auto-correlation has contributions from instrumental noise, $\sigma_{\rm{pip}}$, and the intrinsic variance of the Lyman-$\alpha$ forest flux transmission fluctuations, $\sigma_{\rm{LSS}}$, through 
\begin{equation}\label{eq:sigma_forest}
    \frac{\sigma_q^2}{(\bar{F}C_q(\lambda))^2} = \eta(\lambda)\Tilde{\sigma}_{\rm{pip},q}^2(\lambda) + \sigma_{\rm{LSS}}^2(\lambda),
\end{equation}
where $\Tilde{\sigma}_{\rm{pip},q} = \sigma_{\rm{pip},q}(\lambda)/\bar{F}C_q(\lambda)$, and $\eta(\lambda)$ is a correction factor to account for inaccuracies in the estimation of $\sigma_{\rm{pip}}$. We fix $\eta(\lambda) = 1$ in our analysis following \cite{Ramirez-Perez:2023blu}. The  $\sigma_{\rm{LSS}}$ parameter, which quantifies the noise introduced to our analysis by the small-scale fluctuations in the Lyman-$\alpha$ forest, is fitted iteratively at the same time as the rest-frame continuum $\bar{C}(\lambda_{\rm{RF}})$ and the quasar spectral diversity parameters $a_q$ and $b_q$ \citep{Ramirez-Perez:2023blu}. 

\begin{figure}[!tbp]
    \centering
    \includegraphics[width=\textwidth]{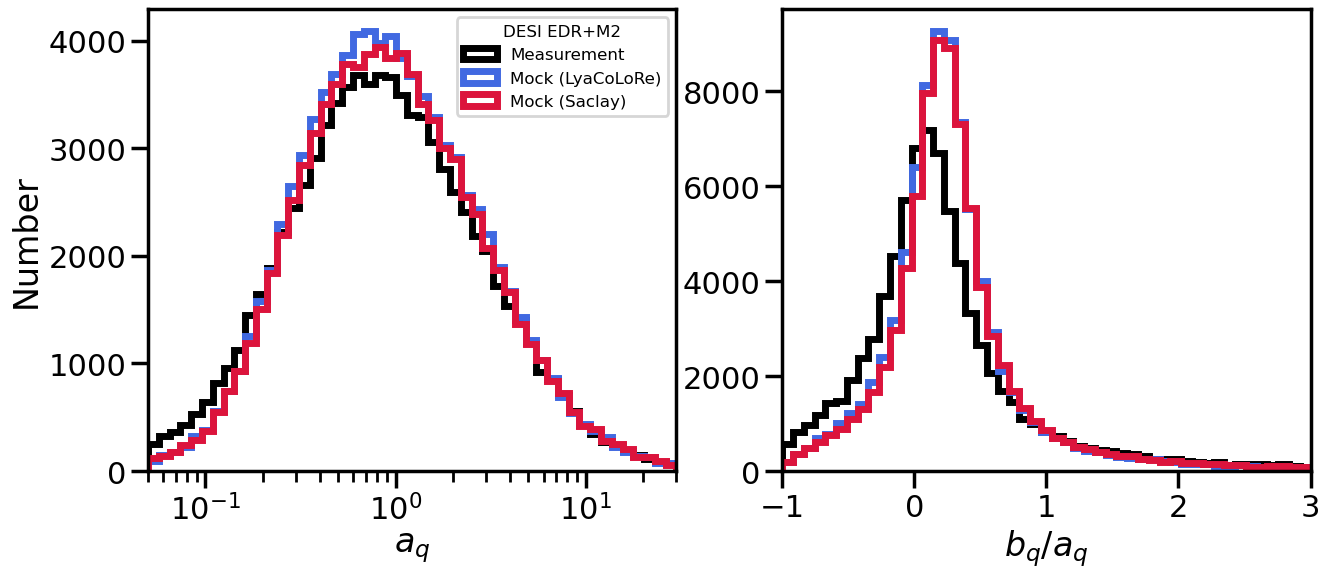}
    \caption{Quasar spectral diversity parameters $a_q$ and $b_q/a_q$ distributions.}
    \label{fig:qsodiversity}
\end{figure}

\Cref{fig:qsodiversity} shows the comparison of the distributions of the quasar spectral diversity parameters $a_q$ and $b_q/a_q$ as measured from data and mocks. We have restricted to those objects whose amplitude parameter is $a_q>0$. The difference between the $b_q/a_q$ distributions of data and mocks suggests that an update in the \texttt{simqso} broken power law model slopes and dispersions might be required to better reflect the quasar spectral diversity observed in the DESI Lyman-$\alpha$ sample.

\begin{figure}[!tbp]
    \centering
    \includegraphics[width=\textwidth]{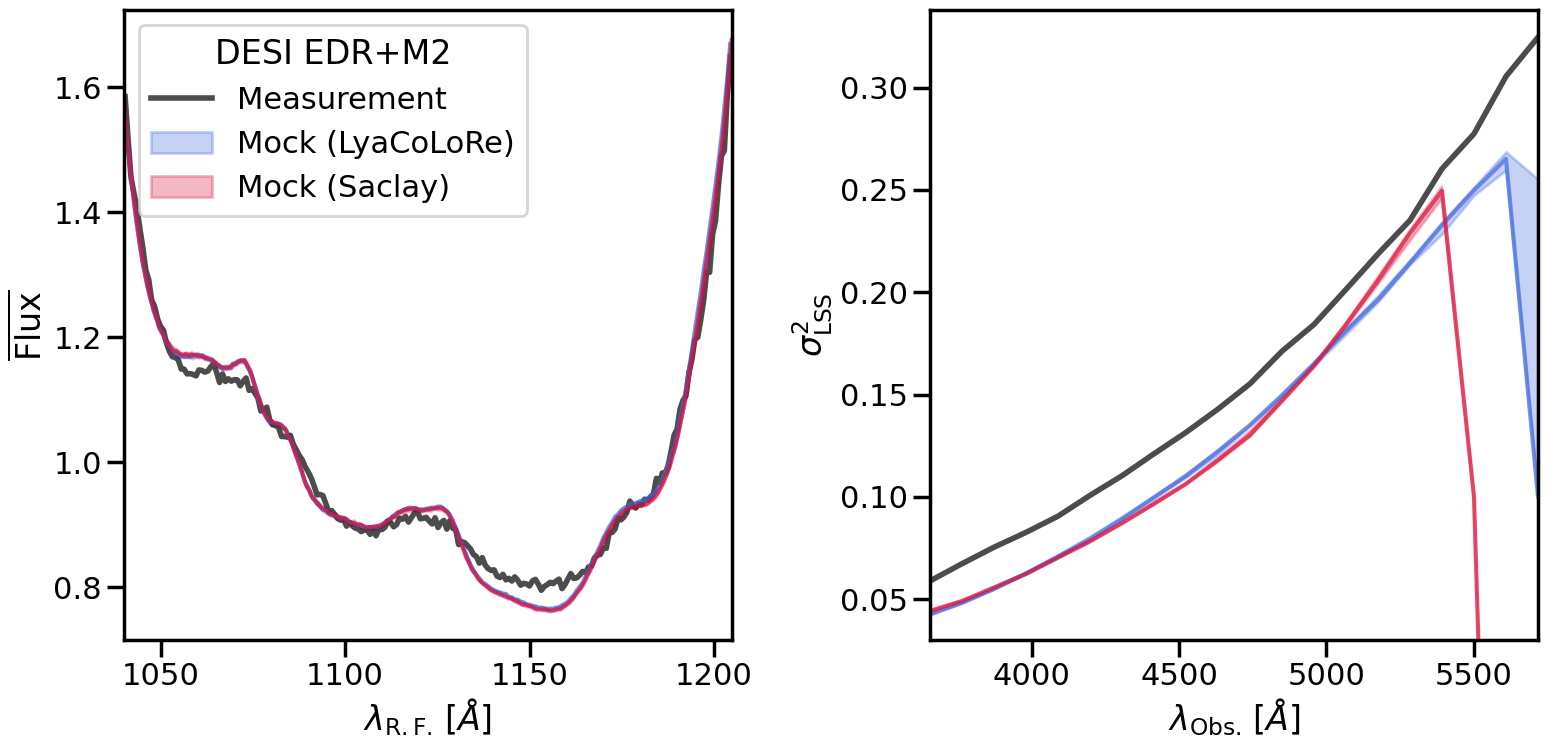}
    \caption{Comparison of the mean continuum at rest frame $C_q(\lambda_{\rm{RF}})$ (left) and the large-scale structure variance $\sigma_{\rm{LSS}}$ (right). Solid line shows median value of 10 realizations of mocks while colored band shows the 1-$\sigma$ percentiles.}
    \label{fig:deltaattribs}
\end{figure}

The left panel of \cref{fig:deltaattribs} shows a comparison of the results of the mean rest-frame continuum $\bar{C}_q(\lambda_{\rm{RF}})$ for data and our mock realizations including contaminants. The mean continuum of each realization shows negligible mock-to-mock variation
and qualitatively agrees with observations, indicating that the emission line model used for this work (described in \cref{subsec:continuum}) is a good approximation. The results for $\sigma_{\rm{LSS}}$ are shown in the right panel of \cref{fig:deltaattribs}. The values obtained by mocks slightly differ from data, which we attribute to the masking in BALs and DLAs, since in mocks we mask the 100\% of these features present in spectra, while in data it depends on the completeness and purity  of the BAL and DLA catalogs, i.e on the performance of the finder algorithms. 
We also see a cutoff of $\sigma_{\rm{LSS}}$ in mocks around $\lambda\approx 5500 \AA$ due to the redshift limit in both \texttt{LyaCoLoRe} (z<3.8) and \texttt{Saclay} (z<3.6) mocks. 

We compute the Lyman-$\alpha$ auto-correlation and Lyman-$\alpha$ -- QSO cross-correlation with the estimators defined in \cref{eq:corr,eq:xcorr}, respectively:
\begin{gather}
    \xi_A = \frac{\sum_{i,j\in A} w_i w_j\delta_i\delta_j}{\sum_{i,j\in A} w_i w_j}, \label{eq:corr} \\
    \xi_A = \frac{\sum_{i,j\in A} w_iw_j\delta_i}{\sum_{i,j\in A} w_i w_j}, \label{eq:xcorr}
\end{gather}
where A is a square bin in separations transverse and parallel to the line of sight with width of 4 Mpc/h. In the auto-correlation estimator, $i$ and $j$ refer to two pixels in the flux-transmission field, while in the cross-correlation estimator $i$ refers to a pixel and $j$ to a quasar. The weights $w^{\rm{Ly\alpha}}_i$ for a flux-transmission field pixel  and $w^{\rm{QSO}}_j$ for a QSO, are respectively defined by
\begin{gather}
    w^{\rm{Ly\alpha}}_i=\frac{1}{\eta(\lambda)\Tilde{\sigma}_{\rm{pip},q}^2(\lambda) + \sigma_{\rm{mod}}^2\sigma_{\rm{LSS}}^2(\lambda)}\left(\frac{1+z_i}{1+2.25}\right)^{\gamma_{\rm{Ly\alpha}}-1}, \label{eq:lya_weights} \\ 
   w^{\rm{QSO}}_j=\left(\frac{1+z_j}{1+2.25}\right)^{\gamma_{\rm{QSO}}-1}, \label{eq:qso_weights}
\end{gather}
where $\gamma_{\rm{Ly\alpha}}=2.9$~\citep{SDSS:2004kjl}, $\gamma_{\rm{QSO}}=1.44$~\citep{duMasdesBourboux:2019zux} and $\sigma_{\rm{mod}}^2$ is an extra parameter introduced to modulate the contribution of the variance of the Lyman-$\alpha$ transmission fluctuations $\sigma_{\rm{LSS}}$. We have fixed the value $\sigma_{\rm{mod}}^2=7.5$ as this value was found to optimize the precision on the results of the correlation functions for the EDR+M2 dataset~\citep{Ramirez-Perez:2023blu}. 

The covariance matrix is estimated by dividing the observed sky region into HEALpix pixels of \path{nside=16} sub-samples and calculating the weighed covariance $C_{AB}$ of two bins $A$ and $B$ by
\begin{equation}\label{eq:covariancematrix}
    C_{AB} = \frac{1}{W_A W_B}\sum_s W_A^s W_B^s \left[ \xi_A^s\xi_B^s - \xi_A\xi_B \right].
\end{equation}
Where $W_A^s$ and $\xi_A^s$ are respectively the summed weight and the measured correlation of the sub-sample $s$ and $W_A = \sum_s W_A^s$. We refer the reader to section 3.2 of \citep{2020duMasDR16}  and 3.5 of \citep{Gordon:2023inprep} for further details on the covariance matrix estimation procedure.

 \begin{figure}[!tbp]
    \centering
    \textbf{Auto-correlation}
    
    \includegraphics[width=\textwidth]{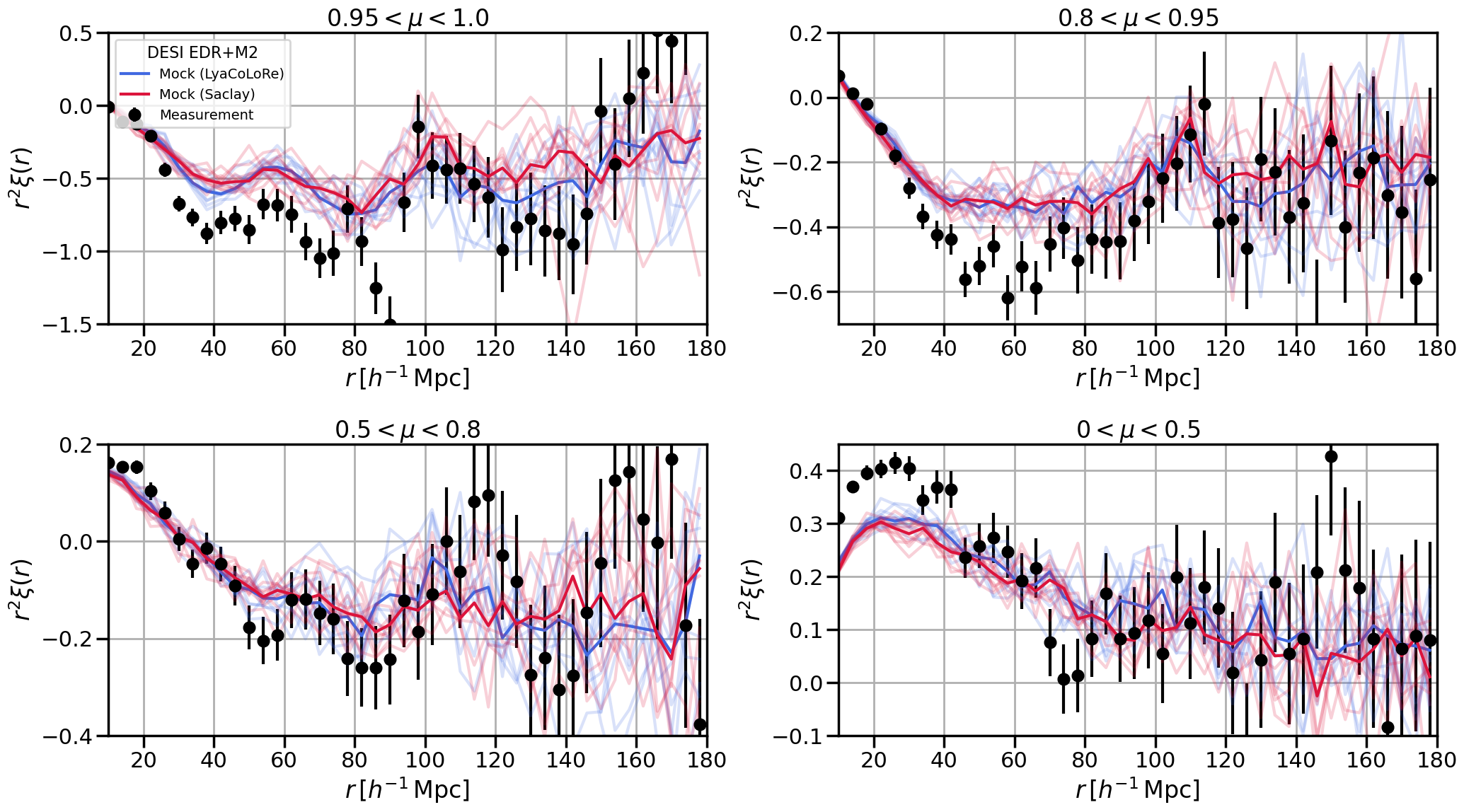}
    
    \textbf{Cross-correlation}
    
    \includegraphics[width=\textwidth]{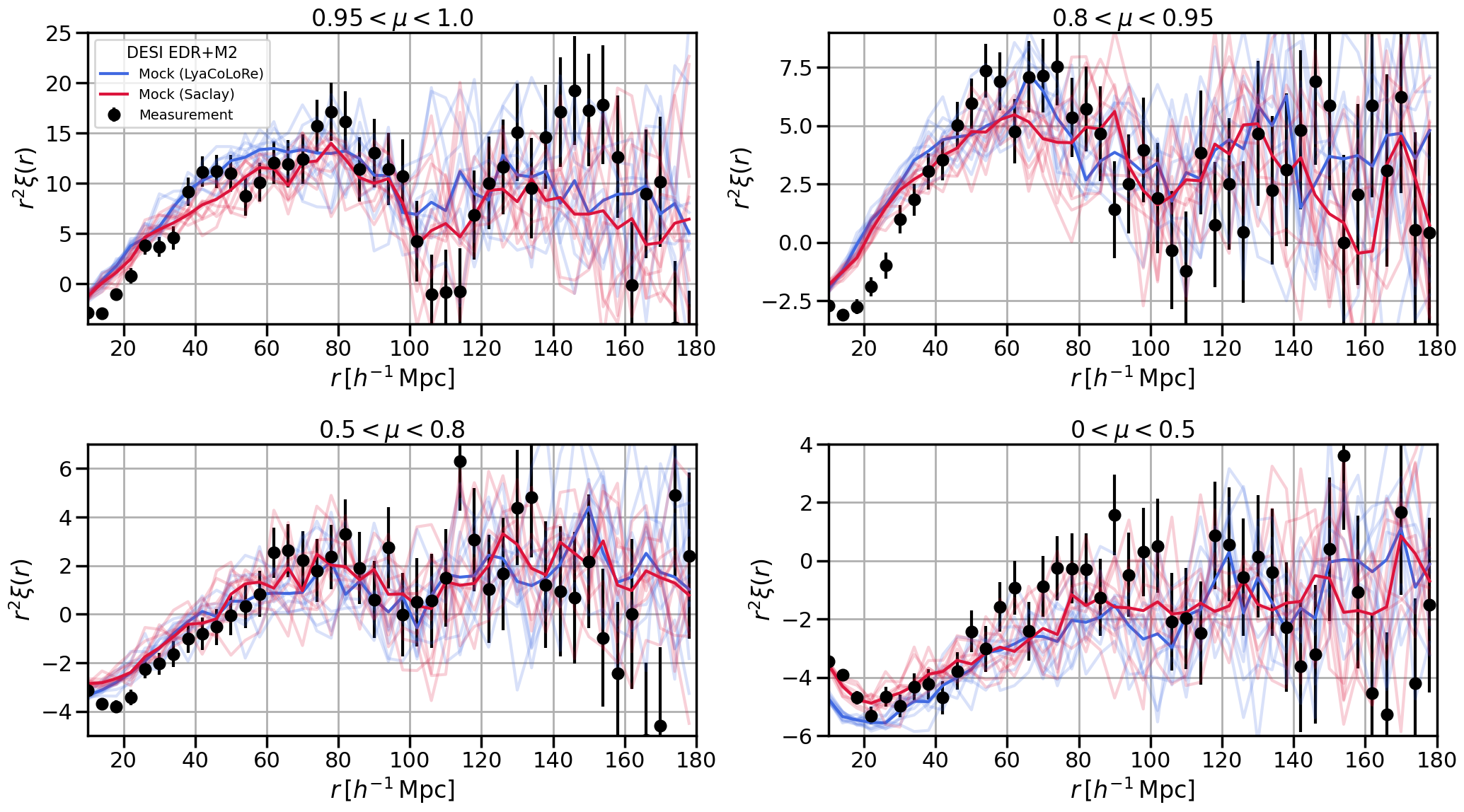}
    \caption{Lyman-$\alpha$ auto (top) and Lyman-$\alpha$ -- QSO cross (bottom) correlation functions multiplied by $r^2$ as averages in four ranges of $\mu=r_\parallel/r$. The transparent red and blue lines show individual correlations of each mock, while the solid line shows the median values. Black dots are the results on observed data as reported by \citep{Gordon:2023inprep}.}
    \label{fig:corrs}
\end{figure} 

\Cref{fig:corrs} shows the Lyman-$\alpha$ auto (top) and Lyman-$\alpha$ -- QSO cross (bottom) correlation functions of the measurements presented by \citep{Gordon:2023inprep} in four ranges of $\mu=r_\parallel/r$. We also show individual correlations of the 10 mock realizations with contaminants for both \texttt{LyaCoLoRe} and \texttt{Saclay} mocks and their median value. We have multiplied the correlations by $r^2$ for better appreciation of the BAO peak. Overall, the computed correlation functions of \texttt{Saclay} and \texttt{LyaCoLoRe} mocks are consistent with observations, however there are some clear differences between mocks and observed data, which we discuss next.

To begin with, we see significant differences at small scales specially at the $0<\mu<0.5$ range of the auto-correlation, and the $0.5<\mu<0.8$ and $0.8<\mu<0.95$ ranges of the cross-correlation, which may be attributed to different reasons. First, the Lyman-$\alpha$ and quasar biases measured on the EDR+M2 observed Lyman-$\alpha$ data, and the target values used to produce the raw mocks are different. The \texttt{LyaCoLoRe} mocks target the results of BOSS DR12, while \texttt{Saclay} mocks target eBOSS DR16. Second, as discussed in \cref{subsec:astrophysical_cont}, systems that go undetected by the DLA finder algorithm, and therefore are not masked, have a significant impact on the shape of the correlation functions, this is not the case for our mocks where we mask all of these features. Third, statistical errors in the measurement of quasar redshifts introduce spurious correlations that mostly affect the shape of the correlation functions in the $0.8<\mu<0.95$ and $0.95<\mu<1$ ranges for both auto and cross correlations~\citep{Youles:2022pgf}, this effect was not included in our mocks. Fourth, instrumental noise has contributions to the shape of the correlation functions more predominantly at small scales of the $0<\mu<0.5$ range. In this regard, there exists the possibility of the model used to introduce instrumental noise not being fully representative of what is measured on data. These possible sources will be studied with mocks of future DESI releases with a higher signal-to-noise ratio. The differences at the $0.95<\mu<1.0$ range at approximately 20 and 60 Mpc/h in the auto-correlation and 110 Mpc/h in the cross-correlation might have some contributions by the value chosen to include metals in our mocks, this will be further discussed during \cref{subsec:astrophysical_EDR}. Finally, note that \texttt{Saclay} mocks have a better reproduction of the cross-correlation at small scales than \texttt{LyaCoLoRe} mocks which is clearly seen at the $0<\mu<0.5$ range, which is expected due to their QSO position sampling method differences discussed in \cref{subsec:rawmock}.

In this work we do not present a best-fit model of the correlations, we just make fits for mocks containing Lyman-$\alpha$ and metal absorptions only as a quality check (see \cref{subsec:astrophysical_EDR}) and use the best-fit values of observed data to validate the use of mocks as a forecast tool (see \cref{subsec:forecast_validation}).
 
\subsection{Astrophysical Contaminants}\label{subsec:astrophysical_EDR}The contaminated mocks include correlated HCDs that follow the \texttt{pyigm} \citep{prochaska2014towards} column density distribution with $\log N_{HI}({\rm cm}^{-2})>17.2$.

The DLA finder algorithm has been found to have purity and efficiency above the 90\% level for systems detected in eBOSS DR16 data with $\log N_{HI}>20.1 \ \rm{cm}^{-2}$ and redshift $z>2.2$ on quasars with
high mean flux $\bar{f_\lambda}>2\times10^{-19}\,\mathrm{W m^{-2} nm^{-1}}$ that generally have high SNR \citep{chabanier2022completed}. However, the performance on DESI data is still under investigation. Therefore, for the purposes of comparing the mock input DLA distribution with the distribution of detected DLAs in the EDR+M2 data, we have conservatively restricted our samples to those systems that fulfill $z_{\rm{DLA}}<z_{\rm{QSO}}$, $z_{\rm{DLA}}>2.6$, $\log N_{HI}>20.5\ \rm{cm}^{-2}$. We have also restricted the redshift of the host QSOs to $2.6<z_{\rm{QSO}}<3.6$ to match the minimum value of the DLA systems in the observed catalog and the maximum value of the \texttt{Saclay} mocks QSOs. Additionally, on data we restrict to DLA systems detected with a confidence larger than 50\% by both the Convolutional Neural Network (CNN) and Gaussian Process (GP) methods of the DLA finder as  was done in ~\citep{Ramirez-Perez:2023blu}. All the mentioned restrictions applied on observed data yield a 34,053 QSOs sample and 2061 detected DLAs. In the case of our mocks we obtain a sample of approximately 35k QSOs and 2,100 DLAs with little mock-to-mock variation.

\begin{figure}[!tbp]
    \centering
    \includegraphics[width=\textwidth]{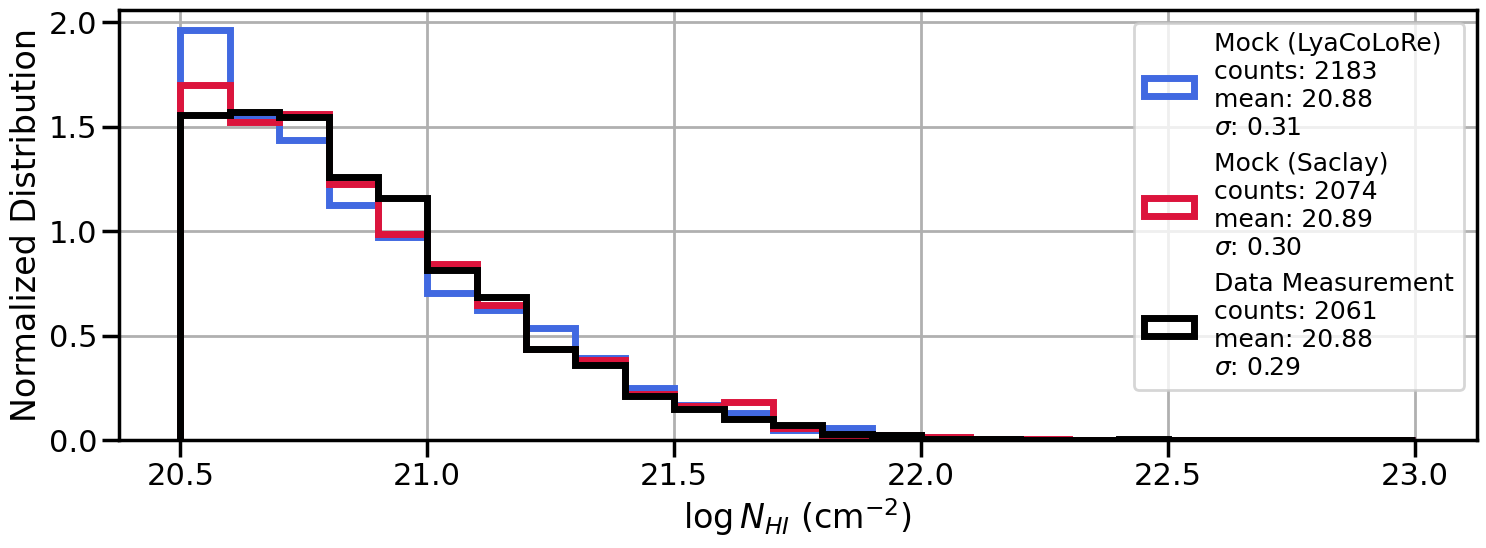}
    \caption{DLA $N_{HI}$ distribution obtained in a EDR+M2 mocks using \texttt{LyaCoLoRe} and \texttt{Saclay} raw mocks as an input with the restrictions $z_{\rm{DLA}}<z_{\rm{QSO}}$, $z_{\rm{DLA}}>2.6$ and $2.6<z_{\rm{QSO}}<3.6$. The distributions were normalized so the cumulative sum is equal to 1.}
    \label{fig:DLA}
\end{figure}

In \Cref{fig:DLA}, we show the resulting distribution of one mock realization of \texttt{LyaCoLoRe} and \texttt{Saclay} mocks, compared with the results of data as obtained by the DLA finder algorithm. We find a discrepancy between mocks and data at $\log N_{HI} = 20.5\ \rm{cm}^{-2}$, which might be due to a lower efficiency of the DLA finder for those column densities. A more exhaustive study of DLAs in our mocks compared to data will be done in future releases of DESI.

\begin{figure}[!tbp]
    \centering
    \includegraphics[width=0.49\textwidth]{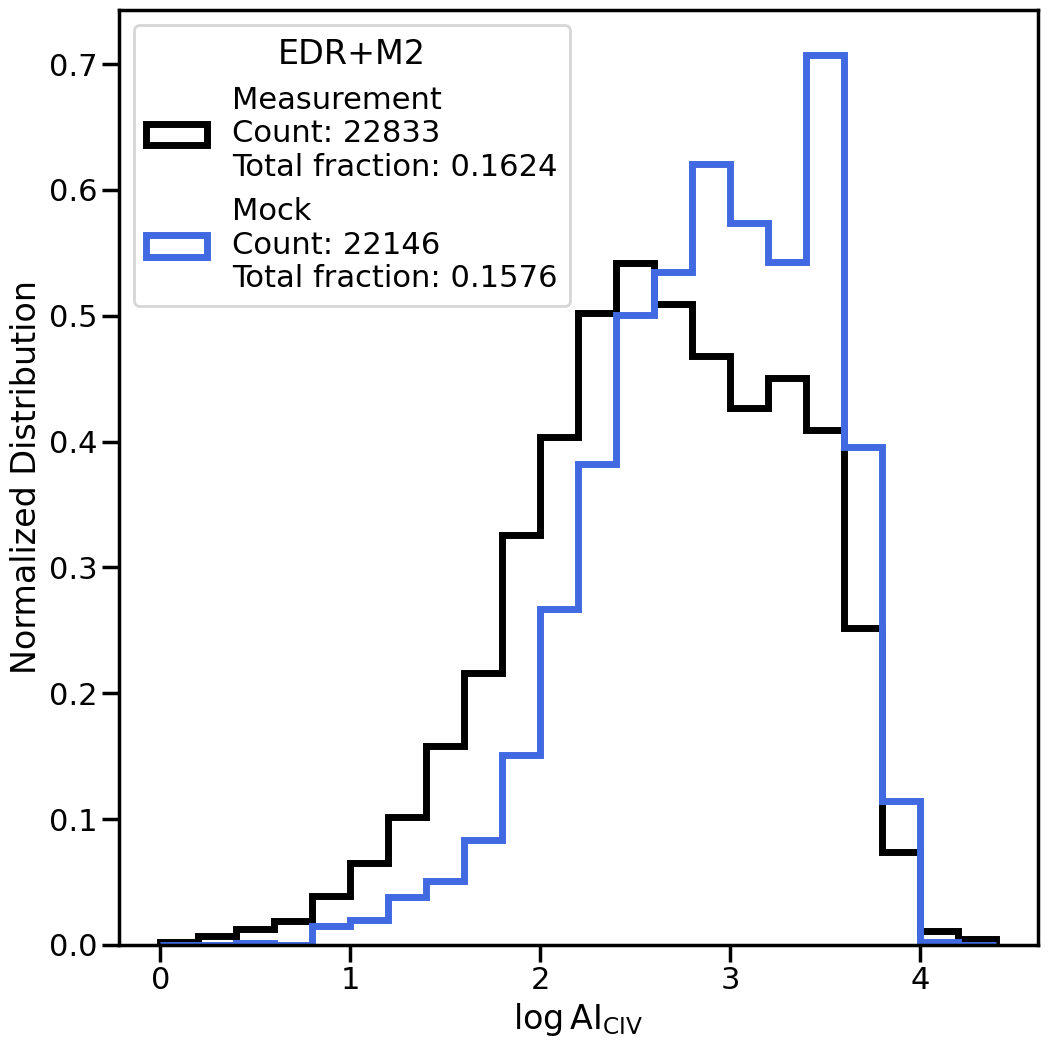}
    \includegraphics[width=0.49\textwidth]{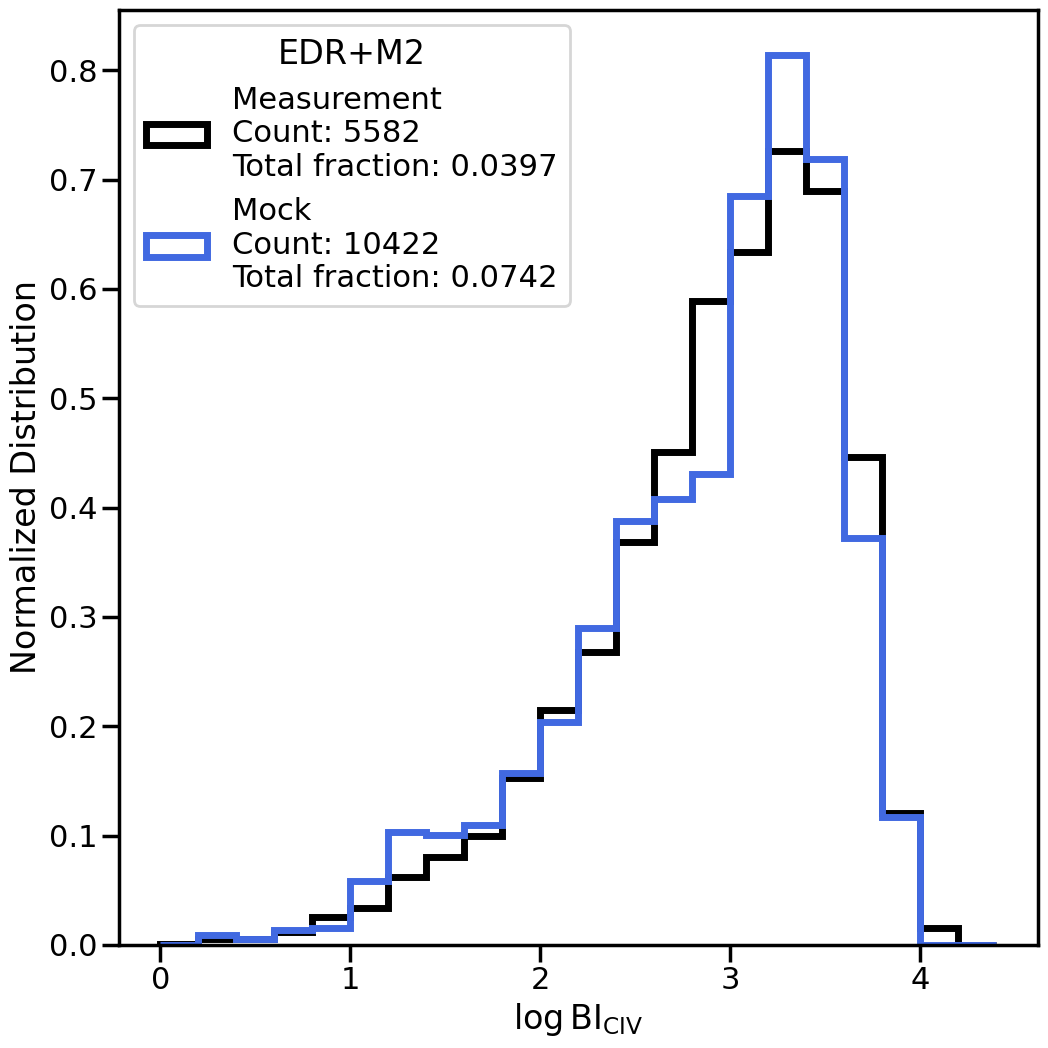}
    \caption{Distribution of $\rm{AI}_{\rm{CIV}}>0$ and $\rm{BI}_{\rm{CIV}}>0$ BAL quasars from DESI EDR+M2 data obtained by the BAL finder algorithm~\citep{Guo:2019bal,Filbert:2023inprep} and one mock realization. We have normalized the distributions so their cumulative sum is 1 by dividing the number of objects in each bin by the total number of objects and by the bin width. In the legend we show the fraction of BAL QSOs  with $\rm{AI}>0$ (left) or $\rm{BI}>0$ (right) divided by the total number of QSO in the sample.}
    \label{fig:BI_AI_distribution}
\end{figure}

The official BAL catalog of the EDR+M2 dataset contains 22.8k BAL quasars with Absorption Index $\rm{AI}>0$ detected by a BAL finder algorithm~\citep{Guo:2019bal,Filbert:2023inprep} when restricting the sample to QSOs within the redshift range $1.8<z<3.6$. For mocks we have used a 16\% probability of a quasar having a BAL in its spectrum, which results in approximately 22.1k $\rm{AI}>0$ BAL quasars for each mock realization. \Cref{fig:BI_AI_distribution} shows Absorption (AI) and Balnicity (BI) indices distributions as obtained by the BAL finder algorithm on data and one mock realization when using the restrictions $\rm{AI}>0$ (left panel) and $\rm{BI}>0$ (right panel). We find agreement between the number of $\rm{AI}>0$ objects on data and mocks, however there is a discrepancy on the shape of the AI distribution, this is discussed in \citep{Filbert:2023inprep} and might be due to the differences between the signal-to-noise ratio of DESI EDR+M2 and the eBOSS DR14 data used to create the BAL templates, we leave the study of the effect of SNR on the distribution shape for future DESI releases with higher statistics than the EDR+M2 sample. While the shape of the BI distribution seems to be in agreement, we have more $\rm{BI}>0$ BAL quasars on mocks than on observations, which we also attribute to the differences between the dataset used to generate the BAL templates and the EDR+M2 data, although this does not affect our results since the analysis pipeline BAL masking criteria is simply $\rm{AI}>0$. Future realizations of DESI Lyman-$\alpha$ mocks might require an update of these templates.

 To compare the effect of metals in our mocks against data we produced additional mocks including only absorptions due to Lyman-$\alpha$, Lyman-$\beta$, Si II(1190), Si II(1193), Si II(1260) and Si III(1207) without DLAs or BALs added. Like in the fully contaminated mocks the metals were included by \texttt{quickquasars} for \texttt{Saclay} mocks and from the transmission file for \texttt{LyaCoLoRe} mocks. We produced 10 realizations for each type of raw mocks. 
 
\begin{figure}[!tbp]
    \centering
    \includegraphics[width=\textwidth]{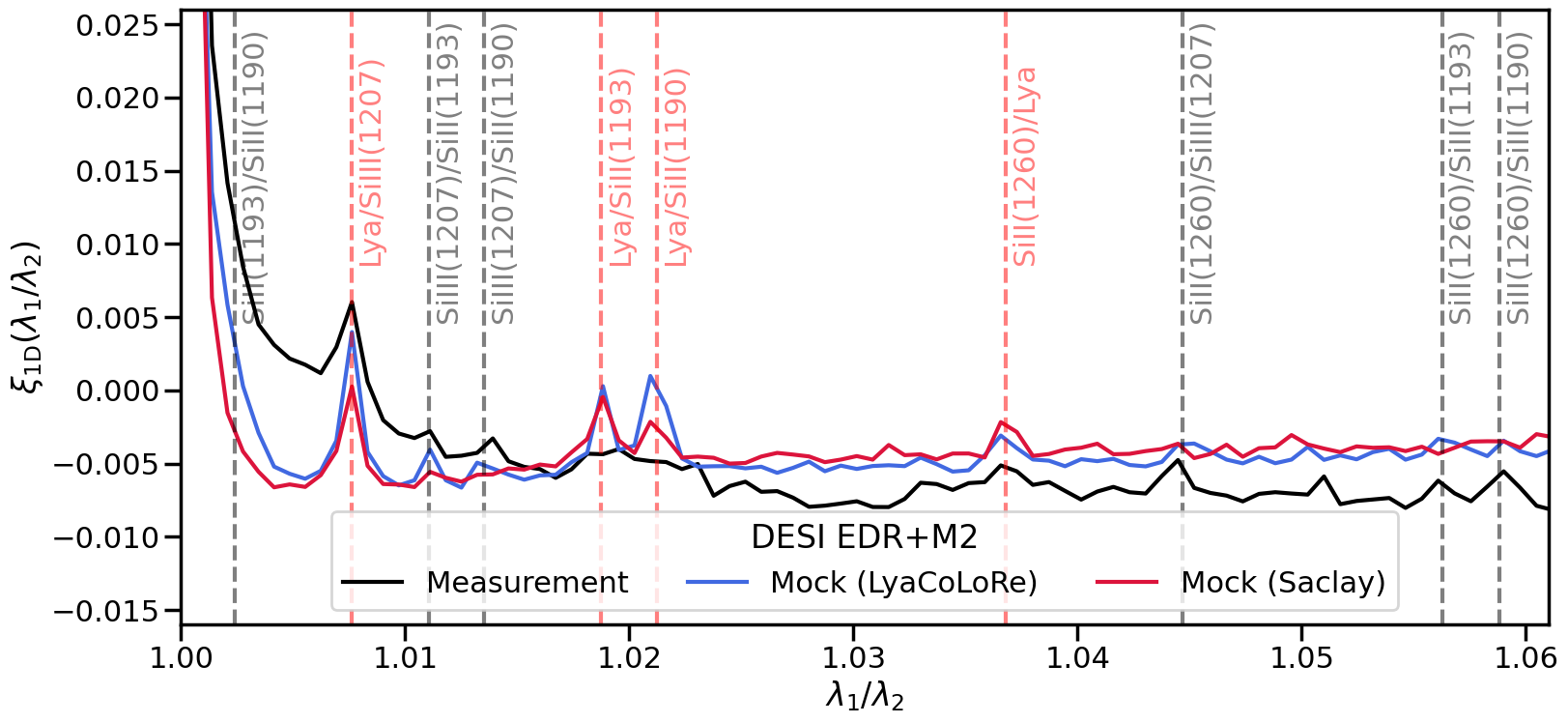}
    \caption{One-dimensional correlation function $\xi_{\rm{1D}}$ of the EDR+M2 measurements and mock realizations including only Lyman-$\alpha$ and metal absorptions. This correlation is expressed as a function of wavelength ratio ($\lambda_1/\lambda_2$) along the same line of sight. We highlight the several Lyman-$\alpha$ -- metal and metal--metal pairs with dashed lines.}
    \label{fig:cf1d}
\end{figure}

First, we are interested on comparing the results of the one-dimensional flux correlation function, given by $\xi_{\rm{1D}}=\langle\delta(\lambda_1)\times \delta(\lambda_2)\rangle$ averaged in bins of $\lambda_1/\lambda_2$. This represents the statistics within individual forests, showing the correlation of $\delta(\lambda)$ as a function of wavelength ratio ($\lambda_1/\lambda_2$) along the same line of sight. It shows prominent peaks due to Lyman-$\alpha$ -- metal and metal--metal correlations.  \Cref{fig:cf1d} shows the comparison of the measured $\xi_{\rm{1D}}$ of one realization for each \texttt{LyaCoLoRe} and \texttt{Saclay} mocks including contamination only due to metals with the results of the observed DESI EDR+M2 data. We highlight with dashed lines the peaks produced by a Lyman-$\alpha$ -- metal transition pairs, along with metal--metal pairs. Observed data includes a prominent peak at $\lambda_1/\lambda_2\approx1.05$ which is not seen in mocks, this is due to a C II(1335) and Si IV(1402) pair, which are not included on the metals we add to our mocks due to these lines requiring further study and a tuning procedure is required to correctly model them. The amplitude of the Lyman-$\alpha$ -- metal pair and the Si II(1260)/Si II(1190) and Si II(1260)/Si II(1193) metal pair peaks in our mocks also show a significant difference compared to observed data. This might be due to the signal to noise ratio differences with observed data which affect the amplitude of these peaks, but might be also due to the coefficients used to include the metals on either type of mock which, as mentioned before, were obtained assuming a linear relation between Lyman-$\alpha$ and metals. Solving this problem requires further study and will be left for future DESI mock realizations.

\begin{figure}[!tbp]
    \centering
    \textbf{Cross-correlation}
    \includegraphics[width=\textwidth]{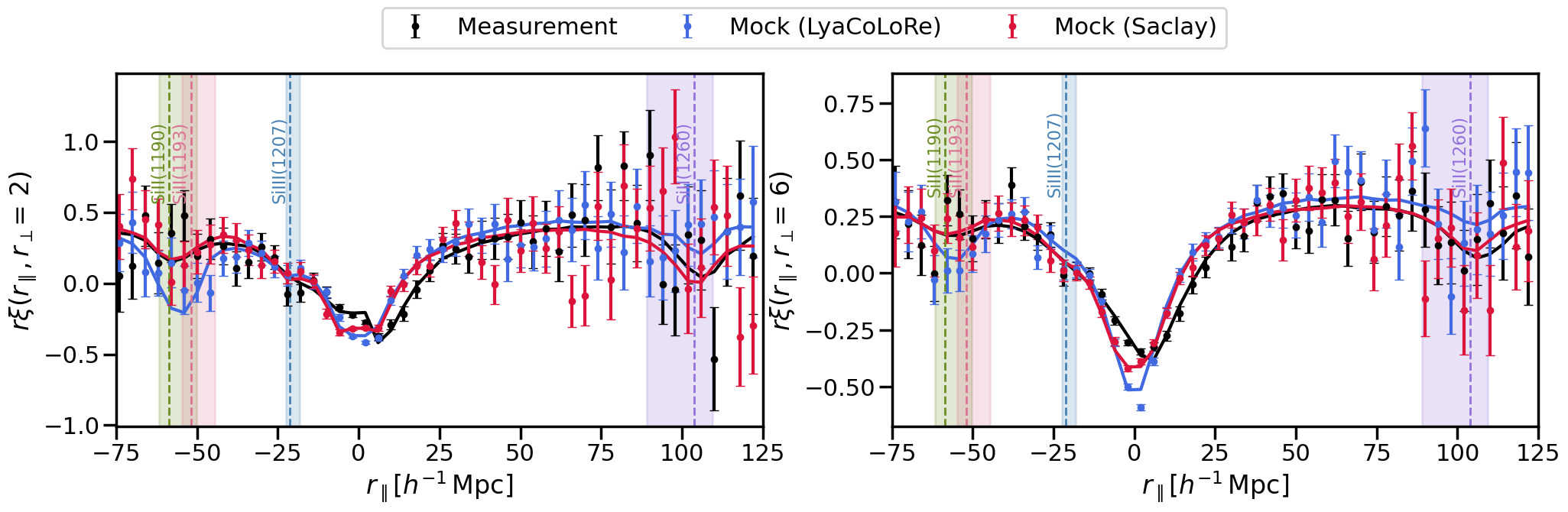}
    \textbf{Auto-correlation}
    \includegraphics[width=\textwidth]{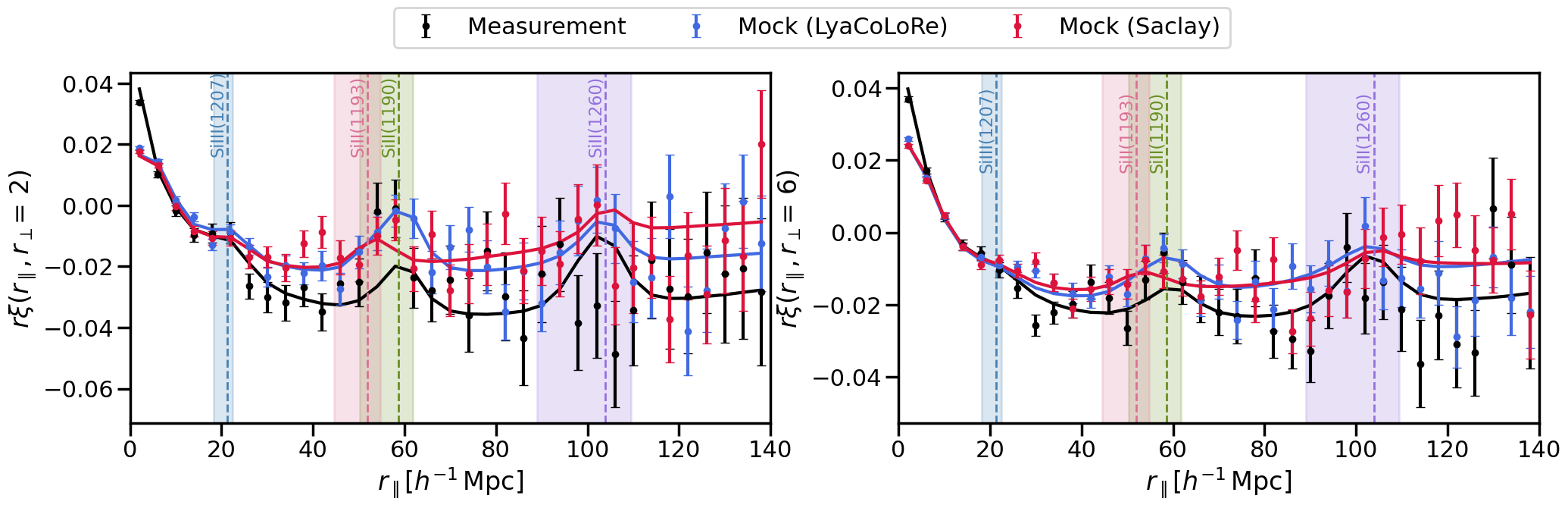}
    \caption{First two $r_\perp$ bins of width 4 Mpc/h centered at $r_\perp=2$ Mpc/h (left) and $r_\perp=6$ Mpc/h (right) of the three-dimensional Lyman-$\alpha$ cross (top) and auto (bottom) correlation functions multiplied by $r$ of the EDR+M2 measurements and mock realizations including only Lyman-$\alpha$ and metal absorptions. Band regions correspond to the scales where each metal distorts the shape of the correlation function $\xi_{\rm{3D}}$ for the wavelength range used in the analysis ($3600\ \AA<\lambda<5772\ \AA$), the dashed vertical lines within the highlighted region is the scale of the metals computed at the effective redshift $z_{eff}=2.37$. Solid lines are the correlation function model as computed using the best-fit value of the velocity bias $b_{\eta,m}$ obtained for these particular mocks, and as reported by the analysis in observed data~\citep{Gordon:2023inprep}.}
    \label{fig:CF_metals}
\end{figure}
 
 Second, we compute the Lyman-$\alpha$ 3D auto-correlation function and perform a best-fit analysis where the velocity bias $b_{\eta,m}=b_m\beta_m/f$ of each metal are measured ($f\approx0.97$ is the growth rate). We use a simple model including Lyman-$\alpha$ and metal correlations, this model is explained in section 4.3 of \citep{Gordon:2023inprep}. We have fixed $\beta_m=0.5$ for all metals following \citep{2017BautistaDR12} due to the fact that $\beta_m$ is poorly determined due to these correlations having significant impact only at small $r_\perp$. 

 The effect of the aforementioned metals is most noticeable in the first two $r_\perp$ bins of the 3D flux correlation functions $\xi_{\rm{3D}}$, which comprise the region closest to the line of sight. In \cref{fig:CF_metals} we show the contributions of Si~II(1190), Si~II(1193), Si~II(1260), and Si~III(1207) to the shape of the Lyman-$\alpha$ cross and auto correlation functions in these $r_\perp$ bins of one realization of the EDR+M2 mocks with metal contamination. These appear as prominent bumps on the cross-correlation at the scale -59, -52, 104 and -21 Mpc/h, respectively, calculated using \cref{eq:metals_rparallel} at the effective redshift $z_{eff}=2.37$. These bumps also appear in the auto-correlation at the absolute value of these same scales. We have multiplied the correlation function by $r$ to clearly show these bumps, but given the low statistics of the EDR+M2 data sample these are not as evident as in the best-fit model, also shown, computed using the velocity bias $b_{\eta,m}$ obtained for these particular mocks.  For comparison with the DESI EDR+M2 data, we also show the measured correlations and fits obtained in \cite{Gordon:2023inprep}. The apparent mismatch in the first $r_\perp$ bin can be attributed to the small statistics in that particular bin, and the fact that the fit is performed using the full correlations.

\begin{figure}[!tbp]
    \centering
     \includegraphics[width=\textwidth]{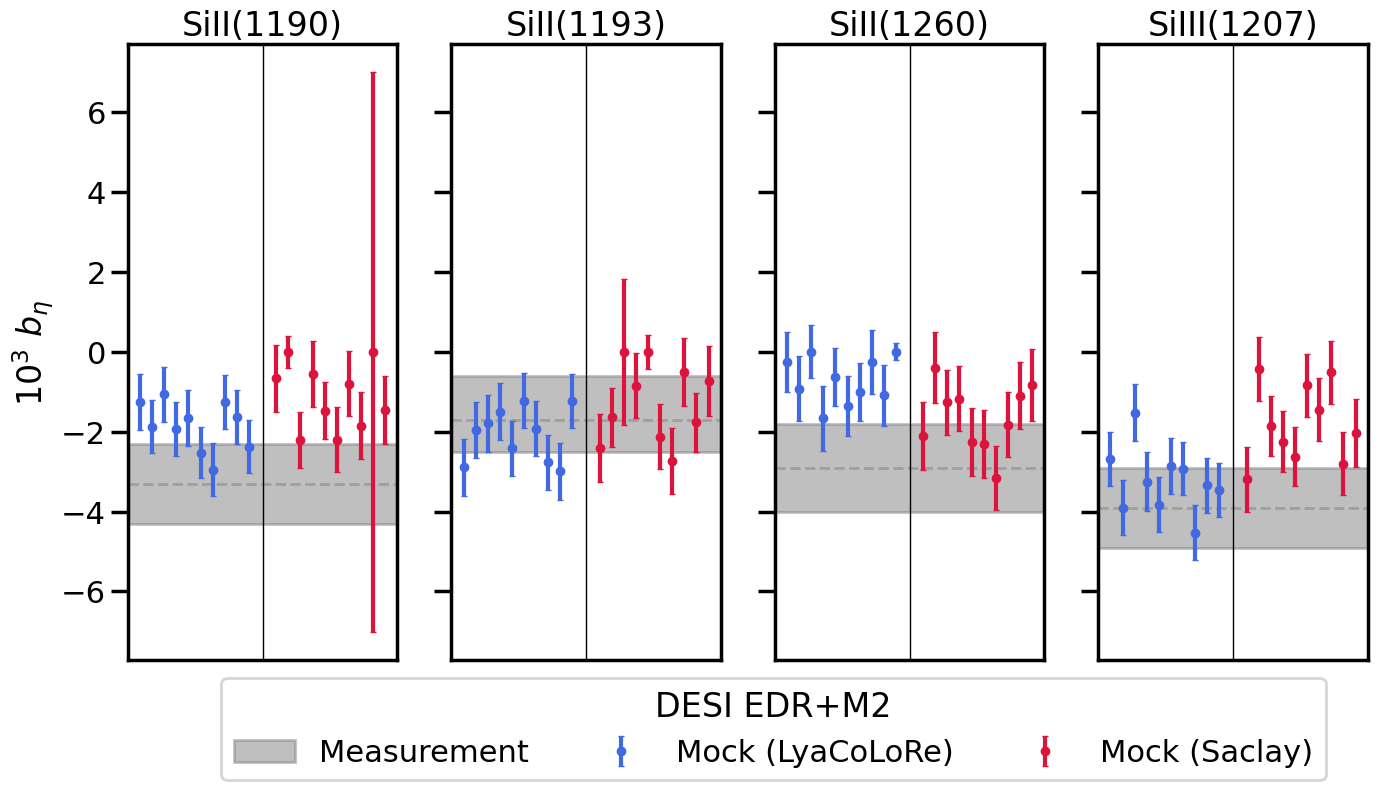}
    \caption{Values of the measured bias $b_\eta$ for four metal transitions in the Lyman-$\alpha$ auto-correlation function. The gray bands show the results of the auto correlation best-fit from DESI EDR+M2~\citep{Gordon:2023inprep} at $10^3b_{\eta,Si\ II(1190)}=-3.3\pm1.0$, $10^3b_{\eta,Si\ II(1193)}=-1.7^{+1.1}_{-0.8}$, $10^3b_{\eta,Si\ II(1260)}=-2.9\pm1.1$, and $10^3b_{\eta,Si\ III(1207)}=-3.9\pm1.0$. The dots are the results of ten realizations of \texttt{LyaCoLoRe} mocks with metals extracted from the transmission files and ten realizations of \texttt{Saclay} mocks with metals added by \texttt{quickquasars}.}
    \label{fig:metal_bias}
\end{figure}

 Finally, in \cref{fig:metal_bias} we present the biases obtained from our 10 \texttt{ LyaCoLoRe} and 10 \texttt{Saclay} mocks contaminated only with metals. Although the overall dispersion of the resulting biases of the included metals is enough to be statistically consistent with the observed data from EDR + M2 for both types of mocks, there is a difference in the mean value of the biases obtained for Si~II(1190), Si~III(1207) bias of \texttt{Saclay} mocks and Si~II(1260) in \texttt{LyaCoLoRe} mocks which contribute to the differences on the correlation functions discussed in \cref{subsec:corrfuncs_EDR}. The results on the metal velocity biases along with the results obtained on the 1D correlation suggests a need for a re-calibration of the relative absorption coefficient of these metals. The possibility of re-tuning the metal coefficients or exploring a non-linear tuning method is left for future DESI data releases.

%% file: Y5mocks.tex
\section{Simulated \texorpdfstring{Lyman-$\alpha$}{Lya} datasets as forecast tool}\label{sec:forecasts}
DESI mocks can be used to forecast the constraining power of the DESI Lyman-$\alpha$ forest dataset, particularly for BAO scale measurement uncertainties, in an alternative and mostly complementary way to a Fisher Forecast. Mock forecast have been useful in a number of situations during the DESI survey preparation stages. For instance, to test the gain DESI would obtain if the target selection methods were able to efficiently select fainter quasars than those considered in the DESI nominal design model, with only one pass, or to determine an optimal observing strategy from the perspective of the Lyman-$\alpha$ studies. 
In this section, we perform a mock based forecast of the constraining power of the full DESI survey. We will do so by using the forecast mode of the \texttt{vega}\footnote{\url{https://github.com/andreicuceu/vega}} package~\citep{Cuceu:2022brl}. This mode starts with the generation of a simulated noiseless correlation function based on a base model given as input. The simulated correlation is then paired with the covariance matrix, obtained from \cref{eq:covariancematrix}, of a mock realization to construct a Gaussian likelihood. Then, for a given correlation function model, with free parameters, to fit this synthetic data, the posterior distribution for the free parameters is sampled using the \texttt{Polychord}\footnote{\url{https://github.com/PolyChord/PolyChordLite}}~\citep{Handley:2015fda,Handley:2015polychord} nested sampler.\footnote{Note that in general the model used to simulate the correlation could differ from the model used to fit it.} Finally, our forecasted uncertainty for each of the free parameters will be given by the 68\% credible region obtained by analyzing the sampling chains using the \texttt{GetDist} package~\citep{Lewis:2019xzd}.

In the next sections, we first apply this methodology to EDR+M2 mocks and compare the forecast uncertainties obtained against the corresponding measurements, as a way to validate the procedure. Then, we perform the full DESI survey forecast and compare the results against a Fisher forecast formalism. 

\subsection{EDR+M2 mock forecast comparison with measurements}\label{subsec:forecast_validation}
We use the Lyman-$\alpha$ auto and Lyman-$\alpha$ -- QSO cross correlations joint fit results obtained by the analysis on observed data (Table 1 in \citep{Gordon:2023inprep}) as our input base model. That is,  we use the same correlation function model as in the observed data analysis which includes non-linear BAO broadening, small-scale corrections with the model from~\citep{Arinyo-i-Prats:2015vqa}, contamination due to DLAs and metals among other features (see section 4 of \citep{Gordon:2023inprep} for a full description of the model). 
The free parameters in this forecast are the bias of Lyman-$\alpha$, HCDs and QSOs ($b_{Ly\alpha}$, $b_{HCD}$  and $b_{QSO}$), the Lyman-$\alpha$ RSD parameter $\beta_{Ly\alpha}$, the velocity bias $b_{\eta,m}$ of each of the metals considered in our mocks, the QSO systematic redshift error shift $\Delta{r}_{\parallel,QSO}$, the statistical redshift errors parameter $\sigma_v$, the quasar radiation effect scale $\xi^{TP}_0$, the instrumental systematic error amplitude $A_{\rm{inst}}$ and the BAO amplitude $A_{\rm{bao}}$. We use flat priors for all free parameters and fit the correlation functions in the $10<r<180$ Mpc/h range. For the covariance matrix, we use one of the \texttt{LyaCoLoRe} mocks realizations which include DLAs, BALs and metals described in \cref{sec:mockvsdata}. Note that we chose to work with \texttt{LyaCoLoRe} mocks over \texttt{Saclay} ones simply because they cover a higher redshift range. \Cref{table:edr_errors} presents the free parameters in the model whose posterior distribution is sampled. We show the central values, that were used to define the base model to perform our forecast, and the uncertainties of such parameters all coming from the measurements in observed data. We also show the uncertainties obtained by our forecast procedure. Although this forecast does not include the BAO scale parameters, $\alpha_\perp$ and $\alpha_\parallel$, since these were fixed to one in the DESI EDR+M2 analysis due to the relatively low statistical power of the data sample~\citep{Gordon:2023inprep}, we can note that the relative difference between the uncertainties obtained by our forecast and the observed data case, defined as $\Delta_\sigma = |1- \sigma_{\rm{Forecast}}/\sigma_{\rm{obs}}|$, are almost all below the 33\% level. We consider that this adequately validates this methodology of using mocks as a forecasting tool, given the differences in the SNR of our mocks compared to data discussed in \cref{sec:mockvsdata} which directly contribute on the results of the covariance matrix used to perform the forecast and therefore the obtained uncertainties. Additionally, our mocks have an approximate representation of small-scale clustering producing a smaller variance than data, which could also have a contribution to the differences obtained. 

\begin{table}
\centering
\caption{Free parameters used for our forecast. We include the central value used as our base model and the 68\% confidence level uncertainties of the auto and cross correlation joint fit as reported by \citep{Gordon:2023inprep} ($\sigma_{\rm{Obs}}$) and the corresponding uncertainties obtained using our forecast procedure on one of the EDR+M2 mocks realization ($\sigma_{\rm{Forecast}}$). Uncertainties with $+$ and $-$ signs indicate the upper and lower value of non-Gaussian posteriors.}
\label{table:edr_errors}
{\footnotesize
\begin{tabular}{lrrr}

\multicolumn{1}{c}{\textbf{Parameter}} & \multicolumn{1}{c}{\textbf{Central value}} & \multicolumn{1}{c}{$\boldsymbol{\sigma_{\rm{Obs}}}$}\citep{Gordon:2023inprep} & \multicolumn{1}{c}{$\boldsymbol{\sigma_{\rm{Forecast}}}$} \\
\hline
\hline
    $b_{Ly\alpha}$                           & $-0.134$ & $0.009$          & $0.006$    \\
    $\beta_{Ly\alpha}$                       & $1.41$   & $+0.12, -0.15$   & $+0.08,-0.1$ \\
    $b_{HCD}$                                & $-0.39$  & $0.009$          & $0.006$    \\
	$10^3 b_{\eta,SiII(1190)}$               & $-2.2$   & $0.8$            & $0.7$    \\
	$10^3 b_{\eta,SiII(1193)}$               & $-0.9$   & $+0.8,-0.3$      & $+0.7,-0.4$ \\
	$10^3 b_{\eta,SiII(1260)}$               & $-2.6$   & $0.9$            & $0.7$    \\
	$10^3 b_{\eta,SiIII(1207)}$              & $-3.4$   & $0.9$            & $0.6$     \\
	$b_{QSO}$                                & $3.41$   & $0.16$           & $0.13$     \\
	$\Delta{r}_{\parallel,QSO} (h^{-1}$Mpc)  & $-2.21$  & $0.18$           & $0.18$     \\
	$\sigma_v (h^{-1}$Mpc)                   & $5.2$    & $0.5$            & $0.4$    \\
	$\xi^{TP}_0$                             & $0.68$   & $0.18$           & $0.14$    \\
	$10^4 A_{\rm{inst}}$                     & $2.4$    & $+0.3,-0.5$      & $+0.3,-0.4$\\
	$A_{\rm{bao}}$                           & $1.17$   & $0.32$           & $0.25$  \\ 
\end{tabular}
}
\end{table}

\subsection{Full DESI survey forecast}
In \cref{subsec:forecast_validation} we discussed the differences between the forecasted uncertainties obtained from our mocks forecasts compared to those measured on data, the results obtained were considered to be favorable given the qualities of our mocks discussed in \cref{sec:mockvsdata}. This allows us to perform a simple but yet informative comparison of the BAO uncertainties forecast obtained by using a mock of the full DESI survey and by the Fisher forecast formalism performed in \citep{DESI:2016fyo} following the procedure described in \citep{Font-Ribera:2013rwa} which uses the method introduced by \citep{McDonald:2006qs} to estimate the uncertainty on the BAO scale measurement using the Lyman-$\alpha$ forest for DESI. In what follows we describe the mocks used for our forecasts and discuss the obtained results.

\begin{figure}
    \centering
    \includegraphics[width=\textwidth]{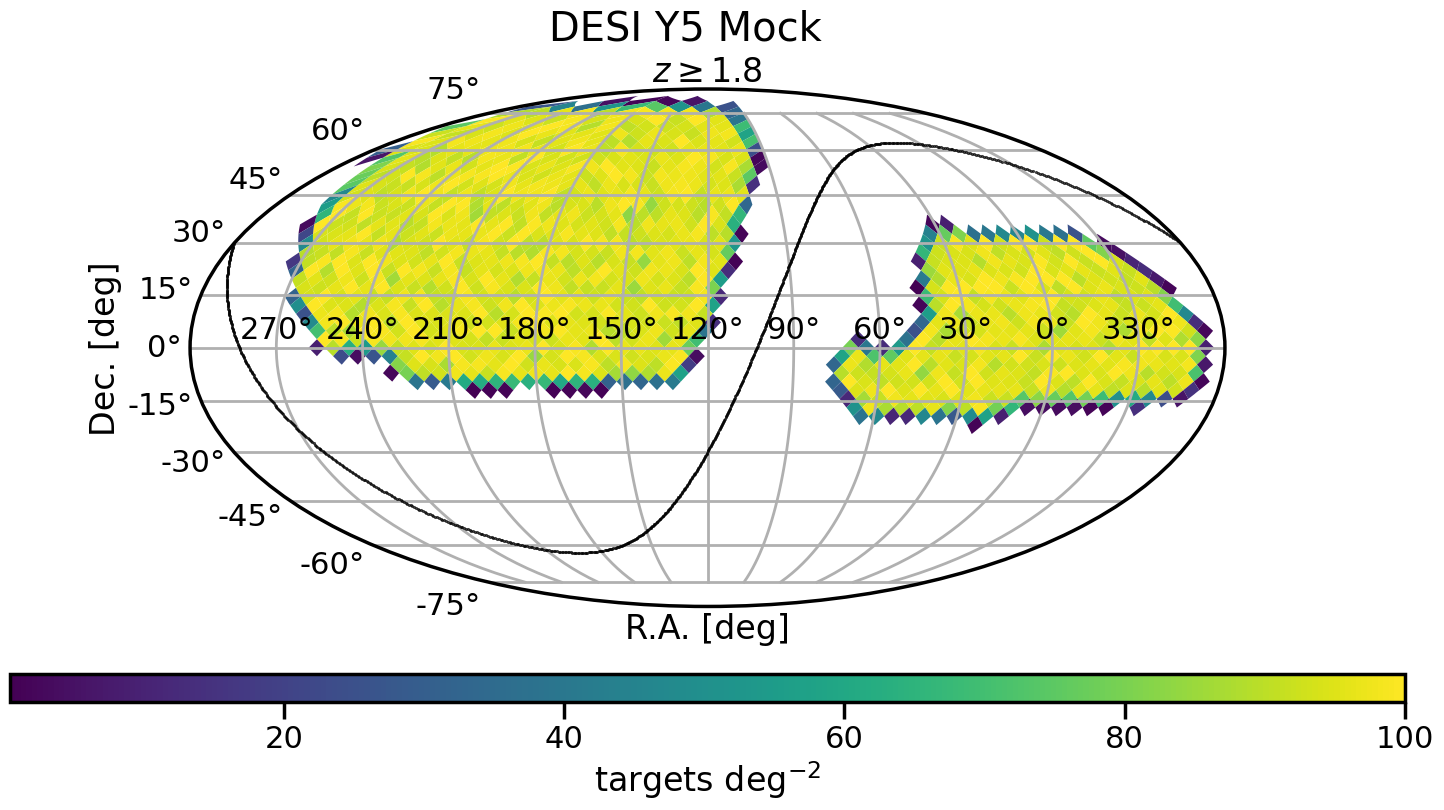}
    \caption{Expected full DESI survey footprint and quasar density  at $z\geq1.8$ divided into \texttt{nside=16} HEALpix pixels.}
    \label{fig:footprintY5}
\end{figure}

We use the methodology previously described to simulate Lyman-$\alpha$ quasar spectra across the 14,000 sq.deg expected to be covered by DESI, assuming a fixed exposure time of 4000 seconds for all targets without galactic extinction added to spectra. For direct comparison with the Fisher forecast we use the same throughput model (black line in \cref{fig:throughput}), redshift-magnitude distribution and object density as expected by the DESI survey nominal design, hereafter referred as the DESI-Y5 DESIMODEL mock. For completeness we also perform a forecast using the redshift and magnitude distributions as observed in EDR+M2, which we will refer simply as the DESI-Y5 mock. 
Doing so results in 1.07 million $z>1.8$ QSOs for the DESI-Y5 DESIMODEL mock, and 1.4 million for DESI-Y5. From these, 698k, and 929k, targets are Lyman-$\alpha$ ($z>2.1$) for each mock, respectively. Figure~\ref{fig:footprintY5} shows the resulting number density of quasars simulated over the DESI footprint, as expected by the observed EDR+M2 redshift and magnitude distributions divided into \path{nside=16} HEALpix pixels.

\begin{table}
\centering
\caption{Base model parameters set in \texttt{vega} to perform the DESI-Y5 DESIMODEL mocks forecasts. The values of $b_{Ly\alpha}$ and $\beta_{Ly\alpha}$ are set to match the first row of Table 1 in \citep{McDonald:2001fe} while the values of $b_{QSO}$ and $\beta_{QSO}$ are set in such way that 
$b_{QSO}=1.2/D(z)$ and $\beta_{QSO}=f/b_{QSO}$, where $D(z)$ and $f=0.966$ are the linear growth factor and rate respectively, computed from the fiducial cosmology (Table 2.2 in \citep{DESI:2016fyo}) at an effective redshift of $z_{\rm{eff}}=2.25$. The value of $D(z)$ is normalized so $D(z=0)=1$. We have set $\sigma_v$ to the value found by DESI EDR+M2 measurement~\citep{Gordon:2023inprep} accounting for the expected statistical redshift errors on observed data.}
\label{table:base_params}
{\footnotesize
\begin{tabular}{ll}
\multicolumn{1}{c}{\textbf{Parameter}} & \multicolumn{1}{c}{\textbf{Value}} \\
\hline
\hline
    $\alpha_\parallel$                       & 1.0       \\ 
    $\alpha_\perp$                           & 1.0       \\
    $b_{Ly\alpha}$                           & $-0.1315$   \\
    $\beta_{Ly\alpha}$                       & 1.58      \\
	$b_{QSO}$                                & 3.092    \\
    $\beta_{QSO}$                            & 0.3123    \\
	$\sigma_v (h^{-1}Mpc)$                   & 5.2      \\
\end{tabular}}
\end{table}

We opted to generate mocks including only Lyman-$\alpha$ absorptions in order to compare to the results of the Fisher forecast formalism presented in \citep{DESI:2016fyo}, which did not include any sort of contaminants in its assumptions. The same forecasts could be performed including contaminants, however this requires further study on the effect of the involved parameter values set on the base model on the resulting uncertainties and goes beyond the scope of this manuscript. 

We measured the Lyman-$\alpha$ auto and Lyman-$\alpha$ -- QSO cross correlations in the Lyman-$\alpha$ and Lyman-$\beta$ regions from the generated mocks following the same analysis procedure as eBOSS DR16~\citep{2020duMasDR16}.\footnote{We do not present the resulting correlations since we are only interested in the resultant constraints.} The resultant covariance matrices are then used for the forecast using \texttt{vega} as in the previous section. For the purpose of this forecast we use a base model  that follows the one used in the Fisher formalism, i.e. it includes a different small-scale correction model than the one used for EDR+M2 mocks (see the first row of Table 1 in \citep{McDonald:2001fe}). However, for parameters that were not specified in the Fisher forecast we used information from  EDR+M2 data. For the DESIMODEL mock this results in the base model parameters presented in \cref{table:base_params}. 
 
In this forecast we set $\alpha_\parallel$ and $\alpha_\perp$ as free parameters along with $b_{Ly\alpha}$, $\beta_{Ly\alpha}$, and $b_{QSO}$ as they are relevant on the correlation function analysis performed on observed data when ignoring contamination from HCDs or metals. Although the Fisher formalism does not include statistical redshift errors we opted to include the measured value of $\sigma_v$ on the EDR+M2 dataset as statistical redshift errors as it is expected to be present through the various DESI data releases, allowing a more realistic uncertainty forecast. 

 The Fisher forecast formalism provides the uncertainties on the Hubble parameter $\nicefrac{\sigma_{Hr_d}}{Hr_d}$ and the angular distance $\nicefrac{\sigma_{\nicefrac{D_A}{r_d}}}{\nicefrac{D_A}{r_d}}$ where $r_d$ is the sound horizon at the drag epoch. In our case, we compute these quantities from the forecasted uncertainty of the BAO parameters $\alpha_\parallel$ and $\alpha_\perp$ and their relationship with $Hr_d$ and  $D_A/r_d$ respectively given by
 \begin{equation}
     \alpha_\parallel = \frac{[(H(z_{\rm{eff}})r_d)^{-1}]}{[(H(z_{\rm{eff}})r_d)^{-1}]_{\text{fid}}},
\end{equation}
\begin{equation}
    \alpha_\perp =  \frac{[D_A(z_{\rm{eff}})/r_d]}{[D_A(z_{\rm{eff}})/r_d]_{\text{fid}}},
\end{equation}
where the "fid" subscript refers to quantities computed assuming a fiducial cosmology, and $z_{\rm{eff}}$ is the effective redshift. Nevertheless, the values of $\nicefrac{\sigma_{Hr_d}}{Hr_d}$ and  $\nicefrac{\sigma_{\nicefrac{D_A}{r_d}}}{\nicefrac{D_A}{r_d}}$ are independent of the chosen fiducial cosmology.
 
To compute the total forecasted uncertainty of the Fisher formalism we use the projections of BAO uncertainties in the redshift range $1.8<z<3.7$ reported in Table 2.7 of \cite{DESI:2016fyo} and use a inverse variance weighting $\sigma=(\sum \sigma_i^{-2})^{-1/2}$, where $i$ corresponds to a redshift bin. This results in $\nicefrac{\sigma_{Hr_d}}{Hr_d}=0.86\%$ for the Hubble parameter and $\nicefrac{\sigma_{\nicefrac{D_A}{r_d}}}{\nicefrac{D_A}{r_d}} = 0.95 \%$ for the angular distance. 

In the case of our forecast using the DESI-Y5 DESIMODEL mock realization we obtain $\nicefrac{\sigma_{Hr_d}}{Hr_d} = 0.71\%$ and $\nicefrac{\sigma_{\nicefrac{D_A}{r_d}}}{\nicefrac{D_A}{r_d}}= 0.87\%$. We attribute the difference partially to the larger wavelength range for the Lyman-$\beta$ forest region analysis of our forecast compared to that of to the Fisher forecast formalism. While the latter uses a wavelength range from $985\ \AA < \lambda < 1200\  \AA$, in the mock analysis we use a wavelength range of  $920\ \AA < \lambda < 1020\  \AA$ and  $1040\ \AA < \lambda < 1200\  \AA$ for the Lyman-$\beta$ and Lyman-$\alpha$ regions, respectively. This was done in order to match the ranges selected by the latest Lyman-$\alpha$ forest in the Lyman-$\beta$ region analysis done in eBOSS DR16.
\begin{figure}
    \centering
    \includegraphics[width=.7\textwidth]{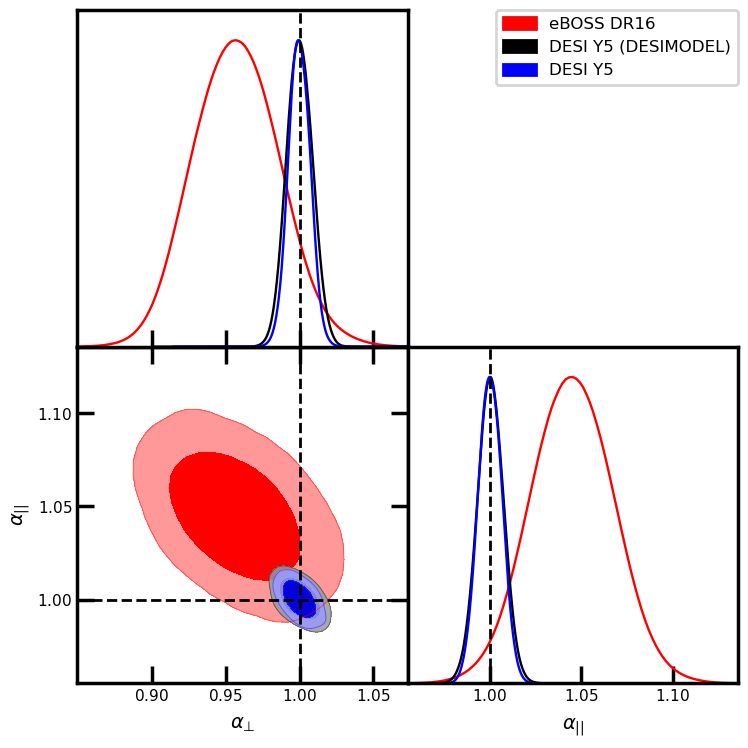}
    \caption{Forecasted uncertainties of $\alpha_\perp$ and $\alpha_\parallel$ as 68\% and 95\% confidence level contour for the DESI-Y5 DESIMODEL mock and for the DESI-Y5 mock. Dashed lines show the values of the fiducial cosmology: $\alpha_\parallel=1$ and $\alpha_\perp=1$. We include in red the results of eBOSS DR16~\citep{2020duMasDR16} as a reference.}
    \label{fig:alphas}
\end{figure}

As for the DESI-Y5 mock, the effective redshift is $z_{\rm{eff}}=2.28$ leading to a small modification of the base model parameters to $b_{QSO}= 3.1195$ and $\beta_{QSO} = 0.3097$. With this configuration we obtain $\nicefrac{\sigma_{Hr_d}}{Hr_d}= 0.64\%$ and $\nicefrac{\sigma_{\nicefrac{D_A}{r_d}}}{\nicefrac{D_A}{r_d}}= 0.73\%$. The different values obtained with respect to the Fisher forecast performed in~\citep{DESI:2023dwi} might be due to the smaller redshift range used in the Fisher forecast, which only include redshifts above $z>2.1$, while we are including quasars down to $z=1.8$ in the Lyman-$\alpha$ -- QSO cross-correlations allowing tighter constraints.

In \cref{fig:alphas} we show the forecasted uncertainty through the 68\% and 95\% credible regions of the BAO parameters $\alpha_\perp$ and $\alpha_\parallel$ for both models and compare with the measurements of DR16~\cite{2020duMasDR16} to highlight the expected constraining power of the completed DESI survey. The results of our DESI-Y5 mock forecasts have a relative difference with the Fisher forecast performed in ~\citep{DESI:2016fyo} of a 17\% for $\nicefrac{\sigma_{Hr_d}}{Hr_d}$ and 8\% for $\nicefrac{\sigma_{\nicefrac{D_A}{r_d}}}{\nicefrac{D_A}{r_d}}$. Regarding the DESI-Y5 mock the results have a relative difference with respect of the Fisher forecast performed in \citep{DESI:2023dwi} of 28\% for $\nicefrac{\sigma_{Hr_d}}{Hr_d}$ and 20\% for $\nicefrac{\sigma_{\nicefrac{D_A}{r_d}}}{\nicefrac{D_A}{r_d}}$.

%% file: appendix_continuum.tex
\section{Emission Lines included in simulated spectra}\label{appendix:emissionlines}
As described in \cref{subsec:continuum} the \texttt{simqso} continuum template generation method relies on a broken power-law model and a Gaussian emission line model defined by the emission line's rest-frame wavelength, equivalent width (and its dispersion), and the Gaussian RMS width ($\sigma$).

\begin{figure}[!tbp]
  \centering
  \includegraphics[width=\textwidth]{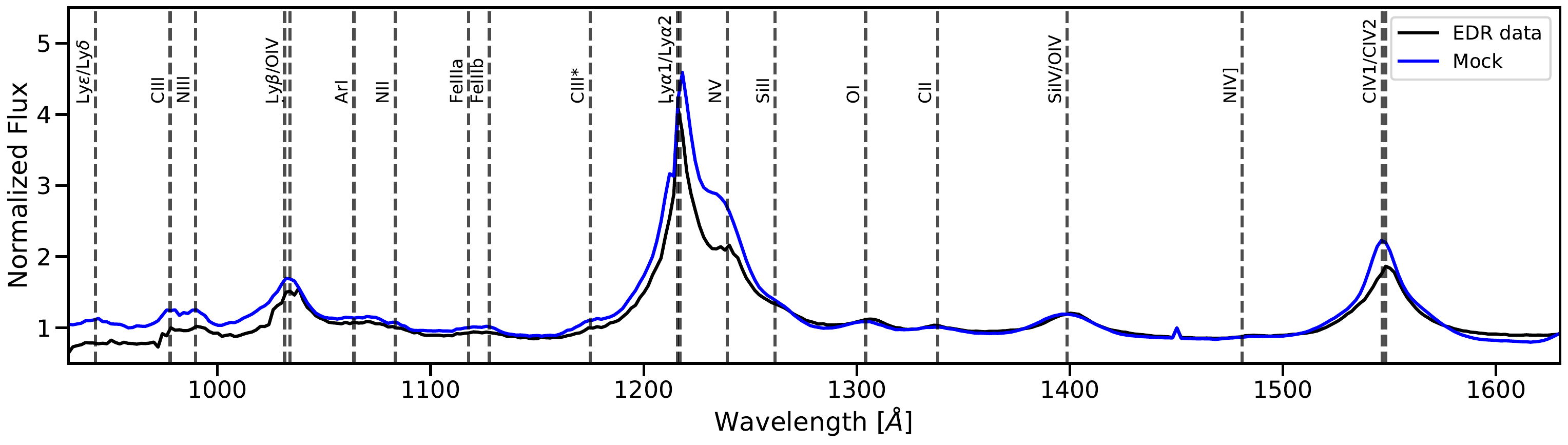}
  \caption{Stack of quasar spectra, normalized at 1450 $\AA$, from DESI EDR+M2 observed data (black) and as obtained from EDR+M2 mocks (blue). The emission lines included in simulated continuum are shown.}
  \label{fig:continuum}
\end{figure}

For the EDR+M2 and DESI-Y5 mocks produced in this work we have defined the emission lines as the combination of two emission line models. One the one hand, we use emission lines model that are within the Lyman-$\alpha$ region of the composite model of the BOSS spectra~\citep{Harris:2016ymq} which was computed from over 102k QSOs in the redshift range of $2.1 \leq z \leq 3.5$ and a rest-frame wavelength range of $800 \AA < \lambda < 3300 \AA$, not including BAL or DLA quasars. On the other hand, outside the Lyman-$\alpha$ region we use the version 7 emission lines model of \texttt{simqso} which includes several datasets of QSO spectra observations on a wider rest-frame wavelength range than the BOSS spectra composite model. We refer the reader to the main \texttt{simqso} repository (given in \cref{footnote:simqso}) for further details on the datasets used to construct this model. The resulting model used for our mocks is shown in \cref{tab:emlinemodel}. We highlight those lines that correspond to the composite model of~\citep{Harris:2016ymq} within the Lyman-$\alpha$ region. The equivalent width of some lines were tuned to resemble the composite model obtained using EDR+M2 spectra. \Cref{fig:continuum} displays a qualitative comparison between composite models of DESI EDR+M2 observed data and a mock realization whose details are given in \cref{sec:mockvsdata}. We can note that mocks contain the main features of QSO continuum.

{\footnotesize
\begin{longtable}[c]{rrr}
\caption{Emission lines in the model used to produce the mocks presented in this work. This model is a combination of the emission lines presented in \citep{Harris:2016ymq} that are within the Lyman-$\alpha$ forest region (highlighted) and the v7 emission line template model of \texttt{simqso}. We only display the wavelength, equivalent width (EW) and Gaussian RMS width ($\sigma$) of the emission lines. We refer the reader to \citep{Harris:2016ymq} and the \texttt{simqso} repository for details about the dispersion of the EW.}
\label{tab:emlinemodel}\\
\textbf{Wavelength {[}$\AA${]}} & \textbf{EW {[}$\AA${]}} & \boldsymbol{$\sigma$}  \textbf{{[}$\AA${]}}\\\hline
\endfirsthead
\multicolumn{3}{c}
{{\bfseries Table \thetable\ continued from previous page}} \\
\textbf{Wavelength {[}$\AA${]}} & \textbf{EW {[}$\AA${]}} & \boldsymbol{$\sigma$}  \textbf{{[}$\AA${]}} \\ \hline
\endhead
629.00    & 6.55   & 5.00   \\
686.00    & 4.16   & 5.00   \\
702.00    & 2.13   & 5.00   \\
773.00    & 11.73  & 10.00  \\
833.00    & 3.14   & 5.00   \\
\rowcolor[HTML]{D9D9D9} 
942.66    & 1.50   & 5.33   \\
\rowcolor[HTML]{D9D9D9} 
977.74    & 1.53   & 4.12   \\
\rowcolor[HTML]{D9D9D9} 
989.73    & 1.50   & 4.14   \\
\rowcolor[HTML]{D9D9D9} 
1,031.48  & 13.53  & 13.15  \\
\rowcolor[HTML]{D9D9D9} 
1,034.07  & 4.26   & 4.84   \\
\rowcolor[HTML]{D9D9D9} 
1,064.01  & 2.90   & 7.66   \\
\rowcolor[HTML]{D9D9D9} 
1,073.53  & 0.71   & 3.51   \\
\rowcolor[HTML]{D9D9D9} 
1,083.31  & 1.32   & 4.11   \\
\rowcolor[HTML]{D9D9D9} 
1,117.85  & 0.76   & 5.26   \\
\rowcolor[HTML]{D9D9D9} 
1,127.55  & 0.46   & 4.11   \\
\rowcolor[HTML]{D9D9D9} 
1,174.91  & 2.49   & 7.68   \\
\rowcolor[HTML]{D9D9D9} 
1,215.85  & 19.08  & 4.85   \\
\rowcolor[HTML]{D9D9D9} 
1,216.94  & 66.63  & 16.56  \\
1,239.28  & 21.21  & 8.75   \\
1,261.67  & 7.42   & 8.88   \\
1,304.12  & 3.81   & 9.16   \\
1,337.91  & 3.28   & 11.07  \\
1,398.69  & 13.43  & 13.58  \\
1,487.87  & 1.50   & 10.51  \\
1,546.55  & 9.56   & 5.60   \\
1,548.25  & 45.35  & 20.49  \\
1,636.23  & 6.86   & 8.23   \\
1,666.39  & 7.30   & 12.87  \\
1,691.30  & 3.68   & 8.98   \\
1,746.34  & 4.13   & 9.34   \\
1,813.20  & 3.61   & 13.90  \\
1,861.66  & 6.00   & 14.41  \\
1,892.70  & 0.93   & 4.49   \\
1,904.12  & 22.38  & 18.20  \\
1,906.87  & 1.86   & 5.28   \\
2,120.00  & 1.70   & 27.00  \\
2,220.00  & 3.00   & 60.00  \\
2,797.86  & 31.42  & 25.03  \\
2,802.95  & 12.80  & 11.40  \\
3,127.70  & 0.86   & 9.38   \\
3,345.39  & 0.35   & 5.50   \\
3,425.66  & 1.22   & 9.09   \\
3,729.66  & 1.56   & 3.32   \\
3,869.77  & 1.38   & 5.31   \\
3,891.03  & 0.08   & 2.02   \\
3,968.43  & 0.45   & 5.32   \\
4,102.73  & 5.05   & 18.62  \\
4,346.42  & 12.62  & 20.32  \\
4,363.85  & 0.46   & 3.10   \\
4,862.68  & 46.21  & 40.44  \\
4,960.36  & 3.50   & 3.85   \\
5,008.22  & 13.23  & 6.04   \\
5,877.41  & 4.94   & 23.45  \\
6,303.05  & 1.15   & 3.14   \\
6,370.46  & 1.36   & 10.18  \\
6,551.06  & 0.43   & 2.21   \\
6,565.00  & 195.00 & 47.00  \\
6,585.64  & 2.02   & 2.56   \\
6,718.85  & 1.65   & 2.09   \\
6,733.72  & 1.49   & 2.54   \\
7,065.67  & 3.06   & 15.23  \\
7,321.27  & 2.52   & 14.26  \\
8,457.50  & 10.70  & 104.20 \\
9,076.80  & 0.80   & 19.00  \\
9,214.00  & 3.50   & 81.40  \\
9,534.40  & 7.00   & 39.90  \\
10,042.30 & 21.10  & 161.20 \\
10,830.00 & 36.00  & 116.00 \\
10,941.00 & 7.00   & 109.00 \\
11,296.40 & 3.30   & 78.80  \\
12,821.30 & 18.40  & 128.30 \\
18,735.50 & 12.70  & 196.10 \\
20,506.80 & 8.60   & 300.50
\end{longtable}}

%% file: appendix_rawmockformat.tex
\section{Raw transmission data model}
\label{appendix:datamodel}
In order for \texttt{quickquasars} to work, the input FITS transmission files should follow the data model of the raw mocks given in \Cref{tab:raw_format}. These files are usually divided by HEALpix pixel following the name convention of \path{transmission-<nside>-<healpix>.fits.gz}, where \texttt{<nside>} is the used nside of HEALpix pixels (\texttt{nside}=16 for the \texttt{LyaCoLoRe}, \texttt{Saclay} and \texttt{Ohio} raw mocks) in NESTED scheme and \texttt{<healpix>} is the index of the HEALpix.

\begin{table}[!htbp]
\centering
\caption{Description of the data format of the transmission files of the raw mocks as required by the input of \texttt{quickquasars}. Optional arguments are not required to be in the file, since these features can be implemented independently by the code.}
\label{tab:raw_format}
{\footnotesize
\begin{tabular}{p{3cm}p{2.3cm}p{8.9cm}}
\multicolumn{1}{c}{\textbf{Name}} & \multicolumn{1}{c}{\textbf{Type}} & \multicolumn{1}{c}{\textbf{Description}}                       \\ \hline \hline
METADATA\footnotemark   & BinTableHDU & Metadata of the mock QSOs (see \cref{tab:tables_meta}).                                \\
WAVELENGTH & ImageHDU    & Wavelength $\lambda$ of the transmitted flux in angstroms.                           \\
F\_LYA     & ImageHDU                          & Lyman-$\alpha$ absorptions transmitted flux. Can also be called TRANSMISSION. \\
F\_LYB     & ImageHDU    & Lyman-$\beta$ absorptions transmitted flux. Optional. \\
F\_METALS  & ImageHDU    & Metal absorptions transmitted flux. Optional.         \\
DLA        & BinTableHDU & Metadata of the correlated DLAs (see \cref{tab:tables_meta}). Optional.               
\end{tabular}
}
\end{table}

\footnotetext{This HDU must include the following information in its header:
\begin{itemize}
    \item HPXPIXEL (int): HEALpix pixel index of the transmission file.
    \item HPXNEST (boolean): True if the NESTED scheme of HEALpix was used.
    \item HPXNSIDE (int): HEALpix \texttt{nside} used for creating the transmission files.
\end{itemize}}

The transmission FITS files include BinTableHDU entries, these should follow the format described in \cref{tab:tables_meta}.
\begin{table}[!htbp]
\centering
\caption{Data format of the BinTableHDU entries of the transmission files.}
\label{tab:tables_meta}
{\footnotesize
\begin{tabular}{p{3cm}p{2.3cm}p{8.9cm}}
\multicolumn{1}{c}{\textbf{Name}} & \multicolumn{1}{c}{\textbf{Format}} & \multicolumn{1}{c}{\textbf{Description}}                       \\ \hline \\
\multicolumn{3}{l}{\textbf{METADATA}}\\ \hline\hline  

RA      &  FLOAT & Right Ascension in degrees.  \\ 
DEC     &  FLOAT & Declination in degrees.  \\ 
Z\_noRSD & FLOAT & Redshift without RSD effects. \\ 
Z       & FLOAT & Redshift. \\ 
MOCKID  & INT & Mock QSO identification number. \\ \hline \\
\multicolumn{3}{l}{\textbf{DLA}} \\ \hline\hline                     
Z\_DLA\_NO\_RSD     &  DOUBLE & DLA redshift without RSD effects.  \\ 
Z\_DLA\_RSD         &  DOUBLE & DLA redshift.  \\ 
N\_HI\_DLA          &  DOUBLE & DLA hydrogen column density. \\ 
MOCKID              &  INT   & Mock quasar identification number. \\
DLAID               &  INT   & Mock DLA identification number. \\
\end{tabular}}
\end{table}

%% file: appendix_output.tex
\section{\texttt{Quickquasars} output files data model}\label{appendix:output}
The final output of \texttt{quickquasars} is a set of FITS files which contain the spectra and relevant information of the quasars divided by HEALpix pixels following the name convention \path{<filename>-<nside>-<healpix>.fits}, where \path{<nside>} and \path{<nside>} is the HEALpix nside and index, respectively; while \path{<filename>} is the type of the stored file, one for each of the following:
\begin{description}
  \item[zbest\footnotemark.] Emulates the results of the DESI pipeline classifier \texttt{redrock}~\citep{redrock2023,2023RedrockQSO}. This includes two HDUS: one with relevant information about the targets, for example: redshift, sky position, and target identification number. The other HDU contains the fibermap.\footnotetext{Constructed as the HDU1 and HDU2 of \url{https://desidatamodel.readthedocs.io/en/latest/DESI_SPECTRO_REDUX/SPECPROD/healpix/SURVEY/PROGRAM/PIXGROUP/PIXNUM/redrock-SURVEY-PROGRAM-PIXNUM.html}}
  
  \item[spectra\footnotemark.] Includes the wavelength, spectrum, and inverse variance for each quasar, divided into tables by each of the R, B, and Z spectrographic bands, and, if requested, the resolution matrix for each band. \footnotetext{Constructed as described in \url{https://desidatamodel.readthedocs.io/en/latest/DESI_SPECTRO_REDUX/SPECPROD/healpix/SURVEY/PROGRAM/PIXGROUP/PIXNUM/spectra-SURVEY-PROGRAM-PIXNUM.html}. By default, and to save space, the resolution matrix is saved only once in the truth files. }
  \item[truth.] This file stores the truth values of the redshift and quasar flux scale before adding astrophysical contributions such as a shift in the redshift due to the Fingers of God effect, flux scale before considering galactic extinction, systematic redshift errors in the pipeline classifier, the random seed, and continuum used to generate each quasar. If applied, this file also contains the information about the resolution matrix, DLAs (column density, redshift) and BALs (redshift, template, balnicity, and absorption indices), for each of the host quasars including their identification number.
\end{description}

The truth FITS files follow the data format given in \cref{tab:truthmeta}, the files contain information about all the generated quasars, except for the DLA and BAL HDUs which only include quasars including these features. Many of the entries in these files are added only if the corresponding feature was used, i.e BALs and DLAs or an option to store the information is given, i.e continuum templates and resolution matrices. \Cref{tab:truthformat} contains the format of each of the BinTableHDU entries of the truth files.

\begin{table}[!htbp]
\centering
\caption{Description of the data format of the truth files stored by \texttt{quickquasars}. Entries marked with an asterisk are included only if the feature was included or if the option to store the information was used.}
\label{tab:truthmeta}
{\footnotesize
\begin{tabular}{p{3cm}p{2.3cm}p{8.9cm}}
\multicolumn{1}{c}{\textbf{Name}} & \multicolumn{1}{c}{\textbf{Type}} & \multicolumn{1}{c}{\textbf{Description}}  \\ \hline \hline
TRUTH          & BinTableHDU & Truth values of the generated quasars. \\
TRUTH\_QSO     & BinTableHDU & Quasars supplemental metadata.\\
DLA\_META*     & BinTableHDU & DLA metadata. \\
BAL\_META*     & BinTableHDU & BAL templates metadata.\\
TRUE\_CONT*    & BinTableHDU & Continuum templates information. \\
B\_RESOLUTION* & ImageHDU    & B-band resolution matrix.   \\    
R\_RESOLUTION* & ImageHDU    & R-band resolution matrix.   \\    
Z\_RESOLUTION* & ImageHDU    & Z-band resolution matrix.   \\    
\end{tabular}
}
\end{table}

{\footnotesize
\begin{longtable}[c]{p{3cm}p{2.3cm}p{8.9cm}}
\caption{Data format of the BinTableHDU entries of the output truth files. Square brackets in some entries mean this entry is an array of the dimension specified inside the bracket, otherwise the entry has a single element.}
\label{tab:truthformat}\\
\multicolumn{1}{c}{\textbf{Name}} & \multicolumn{1}{c}{\textbf{Format}} & \multicolumn{1}{c}{\textbf{Description}} \\ \hline \\
\endfirsthead
\multicolumn{3}{c}{{\bfseries Table \thetable\ continued from previous page}} \\
\multicolumn{1}{c}{\textbf{Name}} & \multicolumn{1}{c}{\textbf{Format}} & \multicolumn{1}{c}{\textbf{Description}} \\ \hline 
\endhead
\multicolumn{3}{l}{\textbf{TRUTH}} \\ \hline\hline                    
    TARGETID       &  INT           & Target identification number. \\
    OBJTYPE        &  STR           & Spectral type. QSO by default. \\
    SUBTYPE        &  STR           & Spectral sub-type (e.g LYA, LRG, ELG). Left blank by default. \\ 
    TEMPLATEID     &  INT           & Continuum template identification number. \\
    SEED           &  FLOAT         & Random seed used to generate spectrum.  \\
    Z              &  DOUBLE        & Redshift.  \\
    MAG            &  FLOAT         & Spectrum normalization magnitude.  \\
    MAGFILTER      &  STR           & Spectrum normalization filter. \\
    FLUX\_G        &  FLOAT         & Flux in G-band. \\
    FLUX\_R        &  FLOAT         & Flux in R-band. \\
    FLUX\_Z        &  FLOAT         & Flux in Z-band. \\
    FLUX\_W1       &  FLOAT         & Flux in WISE W1-band. \\
    FLUX\_W2       &  FLOAT         & Flux in WISE W2-band.  \\
    TRUEZ          &  FLOAT         & Redshift without shifts due to statistical imprecision or systematic shifts. \\
    Z\_INPUT       &  DOUBLE        & Redshift without Fingers of God effect. \\
    DZ\_FOG        &  DOUBLE        & Redshift shift due to Finger of God effect. \\
    DZ\_SYS        &  DOUBLE        & Systematic shift applied to redshift. 0 by default. \\
    DZ\_STAT       &  DOUBLE        & Redshift shift emulating the statistical imprecision of \texttt{redrock}. Present only if applied. \\
    Z\_NORSD       &  FLOAT          & Like Z\_INPUT but without RSD effects. \\
    EXPTIME        &  FLOAT          & Exposure time. Present only if the exposure time was assigned using a distribution.  \\\hline \\
\\\\
    
\multicolumn{3}{l}{\textbf{TRUTH\_QSO}}\\ \hline\hline    
    TARGETID              & INT           & Target identification number.\\
    BAL\_TEMPLATEID       & INT           & BAL template identification number. Set to -1 if the quasar is not a BAL. \\
    DLA                   & BOOLEAN       & True if the quasar hosts a DLA. \\ \\
  \multicolumn{3}{l}{If the \texttt{SIMQSO} continuum template generation method is used:}  \\
    MABS\_1450            & FLOAT         & Rest-frame absolute magnitude at 1450 angstroms. \\
    SLOPES                & FLOAT [5]     & Broken power law continuum slopes. \\
    EMLINES               & FLOAT [3,M\footnote{Denotes the number of emission lines present in the emission line model (73 in our case, see \cref{tab:emlinemodel}).}]   & Emission line parameters used to generate the continuum.\\ \\

  \multicolumn{3}{l}{If the QSO continuum template generation method is used:}  \\
    PCA\_COEFF            & FLOAT [4]       & PCA coefficients used to generate the continuum. \\
    \hline \\
        
\multicolumn{3}{l}{\textbf{DLA\_META}} \\ \hline\hline    
    NHI                   & DOUBLE        & DLA hydrogen column density. \\
    Z\_DLA                & DOUBLE        & DLA redshift.\\
    TARGETID              & INT           & Target identification number. \\
    DLAID                 & INT           & DLA identification number. \\ \hline \\
    
\multicolumn{3}{l}{\textbf{BAL\_META}\footnote{This HDU is constructed to emulate the output of the BAL finder algorithm presented in \citep{Guo:2019bal}. The description of the entries was taken directly from Table 1 of this reference, we refer the reader to this work for further details.}} \\ \hline\hline  
    TARGETID              & INT         &  Target identification number. \\  
    Z                     & FLOAT       &  Redshift. \\ 
    BAL\_PROB             & FLOAT       &  BAL probability. Set to 1 by default. \\
    BAL\_TEMPLATEID       & INT         &  BAL template identification number. \\
    BI\_CIV               & FLOAT       &  C IV Balnicity Index. \\ 
    ERR\_BI\_CIV          & FLOAT       &  C IV Balnicity Index uncertainty. \\ 
    NCIV\_2000            & INT         &  Number of troughs wider that 2000 km/s.\\
    VMIN\_CIV\_2000       & FLOAT [5]   &  Minimum velocity of each absorption trough. \\
    VMAX\_CIV\_2000       & FLOAT [5]   &  Maximum velocity of each absorption trough.\\
    POSMIN\_CIV\_2000     & FLOAT [5]   &  Position of the minimum of each absorption trough. \\
    FMIN\_CIV\_2000       & FLOAT [5]   & Normalized flux density at the minimum of each absorption trough. \\
    AI\_CIV               & FLOAT       & Absorption Index. \\  
    ERR\_AI\_CIV          & FLOAT       & Absorption Index uncertainty. \\  
    NCIV\_450             & INT         & Number of troughs wider that 450 km/s.\\
    VMIN\_CIV\_450        & FLOAT [27]  & Minimum velocity of each absorption trough.\\
    VMAX\_CIV\_450        & FLOAT [27]  & Maximum velocity of each absorption trough. \\
    POSMIN\_CIV\_450      & FLOAT [27]  & Position of the minimum of each absorption trough. \\
    FMIN\_CIV\_450        & FLOAT [27]  & Normalized flux density at the minimum of each absorption trough.\\ \hline \\

\multicolumn{3}{l}{\textbf{TRUE\_CONT}} \\ \hline\hline  
    TARGETID             & INT                     & Target identification number. \\
    TRUE\_CONT           & DOUBLE [W\footnote{Denotes the wavelength array size; it should be the same as the output wavelength arrays of \texttt{quickquasars}.  Default to 3251 for DESI mocks.}] & Continuum template. \\
\end{longtable}}